\documentclass[sigconf, nonacm]{acmart}

\newcommand\vldbpagestyle{plain} 

\usepackage{graphicx}
\usepackage[font=small]{caption}
\usepackage{subcaption}
\usepackage{balance}
\usepackage{xspace}
\usepackage{microtype}
\usepackage{xspace}
\usepackage{colortbl}
\usepackage[T1]{fontenc}
\usepackage{times}
\usepackage[scaled=0.8]{beramono}
\usepackage[LGRgreek]{mathastext}
\usepackage{scalerel}

\definecolor{darkgreen}{rgb}{0.15,0.55,0.15}
\definecolor{mypurple}{HTML}{7030A0}

\newtheorem{example}{Example}

\newcommand{\eat}[1]{}

\newcommand{\stitle}[1]{\vspace{0em}\noindent\textbf{#1}}
\newcommand{\sys}{\textsc{OneProvenance}\xspace}

\usepackage[htt]{hyphenat}
\usepackage{cleveref}

\usepackage{listings}
\definecolor{Keyword}{HTML}{0000ff}
\definecolor{KeywordType}{HTML}{2b91af}
\definecolor{String}{HTML}{a31515}
\lstdefinestyle{tsqlsps}{
	language=SQL,
	showspaces=false,
	basicstyle=\linespread{0.4}\scriptsize\ttfamily,
	numbers=left,
	numberstyle=\tiny,
	xleftmargin=9em,
	linewidth=12cm,
	numbersep=6pt,
	keywordstyle=\color{Keyword},
	stringstyle=\color{String},
	morekeywords={IF,BULK,PROCEDURE},
	classoffset=1, 
	keywordstyle=\color{KeywordType},
	classoffset=2, 
	morekeywords={CleanAndAppendSalesHistory},
	keywordstyle=\color{mypurple},
	classoffset=3, 
	morekeywords={SyncNewSales},
	keywordstyle=\color{darkgreen},
	escapeinside = {(*@}{@*)},
    escapechar=!
}

\newcommand{\syncnewsales}{\textcolor{darkgreen}{SyncNewSales}\xspace}
\newcommand{\cleanandappendsaleshistory}{\textcolor{mypurple}{CleanAndAppendSalesHistory}\xspace}
\newcommand{\trackingsystemversion}{\texttt{@trackingSystemVersion}\xspace}

\newcommand{\stagedsales}{\texttt{StagedSales}\xspace}
\newcommand{\saleshistory}{\texttt{SalesHistory}\xspace}

\newcommand{\xevent}{\texttt{xEvent}\xspace}
\newcommand{\xevents}{\texttt{xEvents}\xspace}
\newcommand{\process}{\texttt{Process}\xspace}

\newcommand{\dataset}{\texttt{Dataset}\xspace}

\newcommand{\asset}{\texttt{Asset}\xspace}

\newcommand{\atlas}{\textsc{Apache Atlas}\xspace}

\newcommand{\openlineage}{\textsc{OpenLineage}\xspace}
\newcommand{\egeria}{\textsc{Egeria}\xspace}

\newcommand{\myhrule}{\rule[.5pt]{\hsize}{.5pt}}
\newcommand{\sstab}{\rule{0pt}{8pt}\\[-2.4ex]}
\newcommand{\mat}[2]{{\begin{tabbing}\hspace{#1}\=\+\kill #2\end{tabbing}}}

\newcommand{\IN}{\mbox{{\bf in}}\ }
\newcommand{\If}{\mbox{\bf if}\ }

\newcommand{\Then}{\mbox{\bf then}\ }

\newcommand{\Else}{\mbox{\bf else}\ }

\newcommand{\Do}{\mbox{\bf do}\ }

\newcommand{\ForEach}{\mbox{\bf for each}\ }

\newcommand{\Return}{\mbox{\bf return}\ }

\newcounter{ccc}
\newcommand{\bcc}{\setcounter{ccc}{1}\theccc.}
\newcommand{\icc}{\addtocounter{ccc}{1}\theccc.}

\makeatletter
\newcommand\xleftrightarrow[2][]{%
	\ext@arrow 9999{\longleftrightarrowfill@}{#1}{#2}}
\newcommand\longleftrightarrowfill@{%
	\arrowfill@\leftarrow\relbar\rightarrow}
\makeatother

\newcommand{\xruntree}{QQTree\xspace}
\newcommand{\activitycollector}{Activity Collector\xspace}
\newcommand{\runtimeinfoextraxtor}{Runtime Information Extractor\xspace}
\newcommand{\lineageextractor}{Provenance Extractor\xspace}
\newcommand{\stitcher}{Stitcher\xspace}
\newcommand{\sqlscriptgenerator}{SQL Script Generator\xspace}

\newcommand{\uploader}{Uploader\xspace}

\newcommand{\sstitle}[1]{\noindent\textbf{#1}}

\usepackage{listings}

\colorlet{punct}{red!60!black}
\definecolor{background}{HTML}{EEEEEE}
\definecolor{delim}{RGB}{20,105,176}
\colorlet{numb}{magenta!60!black}

\usepackage{enumitem}
\newcommand{\takeaways}[1]{\noindent{\bf Takeaways: } #1}
\usepackage[export]{adjustbox}
\usepackage{setspace}

\begin{document}

\title{\textsc{OneProvenance}: Efficient Extraction of Dynamic Coarse-Grained Provenance From Database Query Event Logs [Technical Report]\vspace{-1em}}

\author{Fotis Psallidas, Ashvin Agrawal, Chandru Sugunan$^1$, Khaled Ibrahim, Konstantinos Karanasos$^2$, Jes\'us Camacho-Rodr\'iguez, Avrilia Floratou, Carlo Curino, Raghu Ramakrishnan}

\affiliation{%
	\institution{{\fontsize{9pt}{11pt} \selectfont Microsoft, Snowflake$^1$, Meta$^2$\\first.last@microsoft.com, chandru.sugunan@snowflake.com$^1$, kkaranasos@meta.com$^2$}}
}

\begin{abstract}
	Provenance encodes information that connects datasets, their generation workflows, and associated metadata (e.g., who or when executed a query). As such, it is instrumental for a wide range of critical governance applications (e.g., observability and auditing). 
	Unfortunately, in the context of database systems, extracting coarse-grained provenance is a long-standing problem due to the complexity and sheer volume of database workflows.
	Provenance extraction from query event logs has been recently proposed as favorable because, in principle, can result in meaningful provenance graphs for provenance applications. Current approaches, however, (a)~add substantial overhead to the database and provenance extraction workflows and (b)~extract provenance that is noisy, omits query execution dependencies, and is not rich enough for upstream applications. 
	To address these problems, we introduce \sys: an efficient provenance extraction system from query event logs. \sys addresses the unique challenges of log-based extraction by (a)~identifying query execution dependencies through efficient log analysis, (b) extracting provenance through novel event transformations that account for query dependencies, and (c)~introducing effective filtering optimizations.
	Our thorough experimental analysis shows that \sys can improve  extraction by up to \textasciitilde18X compared to state-of-the-art baselines; our optimizations reduce the extraction noise and optimize performance even further. \sys is deployed at scale by Microsoft Purview and actively supports customer provenance extraction needs ({\small\url{https://bit.ly/3N2JVGF}}).
\end{abstract}

\maketitle

\pagestyle{\vldbpagestyle}

\begingroup
\renewcommand\thefootnote{}\footnote{\noindent $^{1,2}$ Work done while Chandru and Konstantinos were at Microsoft.}\addtocounter{footnote}{-1}\endgroup

\section{Introduction}
\label{s:intro}

Data governance platforms aim to enable organizations to govern (e.g., catalog, overview, secure, analyze, and audit) their data estates. In this direction, Microsoft's governance platform, namely, Purview~\cite{url:purview}, makes a range of governance functionalities readily accessible to customers. (Other such platforms include Collibra~\cite{url:colibra}, Alation~\cite{url:alation}, IBM Infosphere~\cite{url:ibm-infosphere}, or Informatica~\cite{url:informatica}---further highlighting the importance of data governance.) According to several recent business~\cite{url:gartner:governance,url:forrester:governance} and academic~\cite{abadi2022seattlereport} reports, and in-line with our customer feedback,  central to data governance is the ability to capture and use provenance (i.e., a graph encoding connections between inputs and outputs across workflows) and associated metadata (e.g., who or when executed a workflow) from across data systems. Unsurprisingly, since databases play a critical role in data management, capturing provenance from database systems has been one of the most requested feature from Microsoft Purview customers.

Extracting provenance information from database systems is challenging, however, due to the complexity and size of database workflows. Provenance extractors, such as the ones supported by the major governance platforms above, can be classified into \emph{static} and \emph{dynamic} ones. Static provenance extractors access information from database catalogs (e.g., tables, views, and stored procedures), and use static analysis to extract provenance information from them. The main advantage of these extractors is that they can be easily deployed. Unfortunately, however, such extractors can lead to incomplete or incorrect provenance graphs due their inability to monitor the execution of queries (e.g., branches, triggers, or dynamic SQL).  Hence, more recently, dynamic provenance extractors have been introduced to address these limitations. Dynamic provenance extractors operate by listening on events generated by database systems as a side effect of query execution, and extracting provenance from these events. 

Dynamic provenance had also been one of the most highly requested features in Microsoft Purview from customers spanning a wide spectrum of industries (e.g., finance, retail, healthcare, and public services). Designing an efficient and robust dynamic provenance extractor is not straightforward. In particular, based on customer interviews, we find that existing solutions have four main limitations:

\stitle{Design overheads (L1)}: They extract provenance based  on logical or physical plans carried over events generated during query execution (e.g., SAC~\cite{tang:2019:sac}, Spline~\cite{scherbaum2018spline}, or OpenLineage for Spark~\cite{url:openlineage}). As we show in our experiments, this design adds significant overheads (up to \textasciitilde18$\times$) to both query execution and provenance extraction. In real-world scenarios, as highlighted in our customer interviews, these overheads can be prohibitive---both performance- and cost-wise.

\stitle{Limited support for complex scenarios (L2):} They focus on each query in isolation, thus failing to capture important dependencies among queries (e.g., queries executed by a stored procedure). Such dependencies are  ubiquitous and can be complex in practice~\cite{procbench2021}. Hence, resulting provenance graphs are largely incomplete (e.g., no provenance of stored procedures or context on which query triggered another query) and, as such, hard to explore and reason upon.

\stitle{Absence of drill-down/rollup capabilities (L3):} Because existing extractors fail to capture dependencies among queries, customers also pointed out that such extractors do not allow reasoning at various levels of an application. For instance, a user might want to first explore provenance at the stored procedure execution level and then, if needed, drill-down to the provenance of the queries of this stored procedure~\cite{shneiderman1996eyes}. To provide such analysis capabilities, a provenance extractor should be able to aggregate provenance information.

\stitle{Disconnect with consumer applications (L4):} Finally, customers also highlighted that existing extractors fail to provide mechanisms to tailor the resulting provenance information to the needs of the consumer application. For instance, they treat each query as equally important. However, queries issued by system administrators or backup processes are unlikely to be of interest to business users. For such applications, the resulting provenance graphs are often considered noisy and redundant. Similarly, existing extractors lack a rich query runtime metadata model to provide necessary context for upstream applications~\cite{DBLP:conf/cidr/HellersteinSGSA17,mavlyutov2017dependency} (e.g., who executed a query, from which application, or what was the CPU and IO costs). 

To this end, we introduce \sys, a novel dynamic provenance extraction system that addresses the limitations of existing extractors. In particular, 
our system uses a novel extraction design that collects dynamic provenance from low-volume query logs, without relying on plans 
(\textbf{L1}). As such, \sys avoids excessive overheads on database execution and provenance extraction to the extent that \sys extracts provenance from even sizeable transactional workloads---extending coverage beyond the traditional focus on analytical and ETL workloads.

\sys tackles complex scenarios by identifying query dependencies through efficient query log analysis (\textbf{L2}) and providing drill-down/rollup capabilities over these query dependencies (\textbf{L3}). More specifically, query dependencies are encoded in a novel tree data structure, that we refer to as \xruntree. Provenance is then aggregated based on parent-child relationships of \xruntree{}s.

To better accommodate application-specific requirements (\textbf{L4}), we introduced filtering techniques to eliminate redundant provenance information in various points of the provenance extraction workflow. Moreover, \sys employs an extensible query runtime metadata model that allows capturing application-specific metadata in the provenance graph. Importantly, our data model is in compliance with open standards (\atlas~\cite{url:atlas}) for better interoperability with existing metadata management systems. 

To summarize, our key contributions in this paper include:

\begin{itemize}[leftmargin=*,topsep=.3em,itemsep=0mm,partopsep=-.5em]
	\item A data model encoding dynamic provenance, metadata, and query dependencies---while complying with open standards (\Cref{s:bg}).
	\item An efficient and extensible dynamic provenance extractor that overcomes several limitations of existing extractors (\Cref{s:onep}).
	\item Filtering techniques to optimize the extraction process depending on application requirements (\Cref{s:opts}).
	\item The integration of \sys with Purview (\Cref{s:integration}).
	\item A thorough experimental analysis across workflow types (from transactional to analytical ones) highlighting the performance of \sys (end-to-end, in individual components, and in comparison with state-of-the-art extraction techniques), and the benefits from our proposed optimizations (\Cref{s:exps}).
\end{itemize}

\section{Running Example}
\label{s:running}

\begin{figure}[t]
	\centering
	\begin{lstlisting}[style=tsqlsps]
CREATE PROCEDURE CleanAndAppendSalesHistory
   @trackingSystemVersion int
AS
BEGIN
    IF @trackingSystemVersion = 1
      BEGIN
        INSERT SalesHistory
        SELECT c.CustomerId, c.Region,
               r.Rate * c.Amount AS Amount
        FROM   StagedSales c JOIN
               ConversionRate r ON c.Region = r.Region
      END
    ELSE
      BEGIN
        INSERT SalesHistory SELECT * FROM StagedSales
      END
END !\vspace{.5em}!
CREATE PROCEDURE SyncNewSales
   @trackingSystemVersion int
AS
BEGIN
    IF EXISTS(SELECT * FROM INFORMATION_SCHEMA.TABLES
              WHERE TABLE_NAME='StagedSales')
       DELETE FROM TABLE StagedSales;
    BULK INSERT StagedSales FROM 'newSales.csv';
    EXECUTE CleanAndAppendSalesHistory
            @trackingSystemVersion;
END !\vspace{.5em}!
EXECUTE SyncNewSales 2;
\end{lstlisting}
	\vspace{-3mm}
	\caption{Workflow of our running example.}
	\label{f:ex-workflow}
	\vspace{-1.2em}
\end{figure}

To better highlight key points in our discussion, we use the following running example throughout the rest of the paper:

Consider the T-SQL script in~\Cref{f:ex-workflow}. The stored procedure \syncnewsales (lines $18$-$28$) is populating the table \stagedsales (deleting its previous contents, if it already exists) using an external CSV file. Then, it calls the stored procedure \cleanandappendsaleshistory (lines $1$-$17$) that is responsible for appending new data to the \saleshistory table. Note that the query that populates \saleshistory varies depending on the value of the parameter \trackingsystemversion. Hence, provenance information and runtime metadata for these two stored procedures may change across their runs. 

Using this example, we will show how we can extract dynamic provenance, identify dependencies (e.g., across executions of a stored procedure or queries that are part of a stored procedure), attach metadata to nodes of the provenance graph (e.g., who and from what server and application executed the stored procedure or how much CPU time and I/O was taken during execution of a stored procedure), and aggregate provenance across queries.

\section{Data model and problem statement}
\label{s:bg}

\newcommand{\qlog}{$Q_{log}$}

Our overall goal is to extract semantically-rich, coarse-grained provenance information from executed queries. In this section, we discuss our provenance model and associated problem statement. 

\subsection{Provenance Model}
\label{s:bg:datamodel}

At a high-level, we model coarse-grained provenance as a hypergraph that captures the relationships between datasets (e.g., tables or columns) and processes (e.g., queries). We now define the building blocks for modeling such a graph (i.e., our provenance model). 

\begin{figure}[t]
	\centering
	\includegraphics[width=.8\columnwidth]{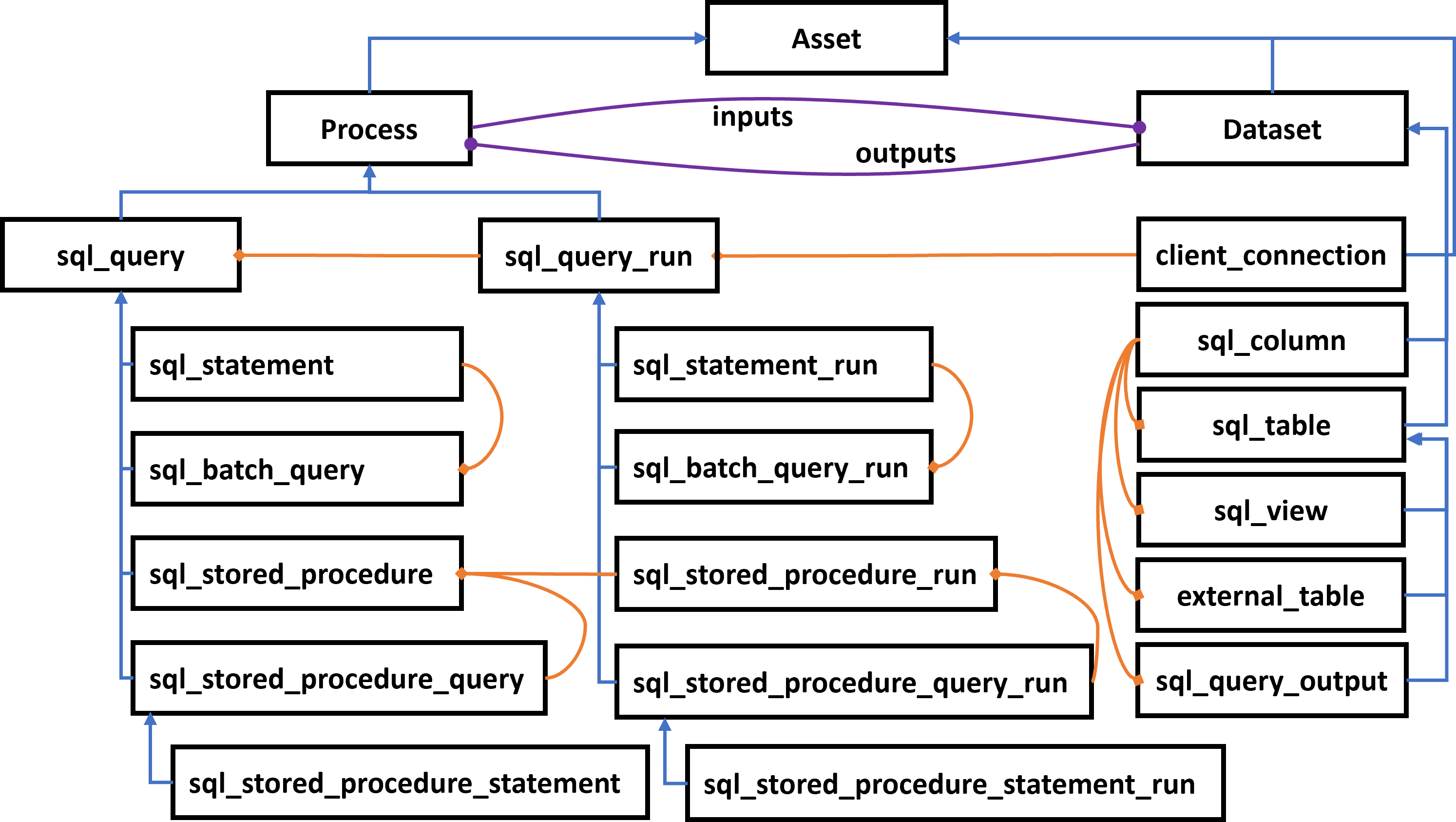}
	\vspace{-.8em}
	\caption{\sys Data Model. Blue lines with arrows correspond to inheritance relationships, orange lines with diamonds correspond to containment relationships, and purple lines with circles highlight provenance relationships. (We omit attributes per type for brevity.)}
	\label{f:datamodel}
	\vspace{-1.5em}
\end{figure}

Since we aim to comply with open standards (recall~\Cref{s:intro}), our provenance model is built on top on the, heavily extensible, \atlas type system. More specifically, \atlas introduces the generic entity types \process and \dataset. (Both types are derived from the generic \asset type of \atlas, as shown in~\Cref{f:datamodel}.) Metadata on processes and datasets can be introduced as attributes (e.g., a dataset may have a name, size, and id) or through relationships (e.g., a client connection invokes a process). \atlas supports inheritance and containment relationships that are relevant to our work as we discuss below. Provenance is encoded as special relationships between \process and \dataset entities denoting the input/output datasets of a process. (Finally, note that \atlas is only one target type system for provenance information. Our model can also be compiled to other standards including OpenLineage~\cite{url:openlineage} or W3C PROV-DM~\cite{missier2013w3c}).

With the background of \atlas in place we can now dive into our model which is depicted in~\Cref{f:datamodel}. 

\sstitle{Datasets.}  In line with prior work on coarse-grained provenance~\cite{tang:2019:sac,scherbaum2018spline}, datasets in our model include relations (such as tables, views, external tables, and query outputs) and their associated columns.

\sstitle{Processes.}  In our model, processes can be either \textit{queries} or \textit{query runs}. The former is traditionally the target of static provenance extractors, and the latter the target of dynamic ones. Queries and their runs are then sub-typed through inheritance relationships. As shown in Figure~\ref{f:datamodel}, in our current model, we encode ad-hoc statements, batches (encoded as a series of statements), queries that are part of stored procedures, and stored procedures. For every such static query type, we introduce its dynamic type by subtyping on query run.

In addition, we use attributes to encode metadata for processes. In particular, for queries we track the query text, while for query runs we track the user that executed the query; CPU time; duration; rows inserted, updated, deleted, and returned. Finally, a client connection is attached to each query run through a containment relationship to encode from what application and server the query run was invoked.

\sstitle{Query dependencies.} To support complex scenarios and enable drill-down and rollup capabilities (addressing L2-3 from~\Cref{s:intro}), we introduce two types of dependencies: (1) query runs spawned by parent query runs and (2) runs of a query. The former allows us to encode what query runs have been triggered by other queries (e.g., queries executed as part of executing a stored procedure) and vice versa (e.g., what stored procedures a query has been part of). The latter allows us to track the runs of a particular query (e.g., the runs of a stored procedure). Both types of dependencies are encoded through containment relationships in our model (see also~\Cref{f:datamodel}).

\sstitle{Provenance.} Following our conceptual model, we model provenance as a hypergraph $P$ connecting inputs with outputs across a workflow. Logically, each edge $i \xleftrightarrow{p} o$ maps input $i$ to output $o$, derived from $i$, through process $p$. As we discussed above, in our case, such processes are query runs and their static counterparts, while input and output datasets are relations and columns.

As discussed in~\Cref{s:intro}, our main contribution to address \textbf{L3} is to aggregate provenance through the query dependencies of our model. We define the \textit{provenance of a query run} as the set union of the provenance of the query runs that were executed as a result of executing that query (i.e., for query run $Q_r$ we define its output $O(Q_r)$ as $O(Q_r)=\bigcup_{Q_r\xleftarrow{}Q'_r} O(Q'_r)$, where $Q_r\xleftarrow{}Q'_r$ denotes the set of query runs $Q'_r$ triggered by $Q_r$; similarly for inputs). Furthermore, we define the \textit{provenance of a query} as the set union of the provenance of its query runs (i.e., for query $Q_s$ we define its output $O(Q_s)$ as $O(Q_s)=\bigcup_{Q_s\xleftarrow{}Q_r} O(Q_r)$, where $Q_s\xleftarrow{}Q_r$ denotes the set of query runs $Q_r$ for the query $Q_s$; similarly for inputs). Note that provenance is defined recursively for both definitions. The base case is the provenance of individual statements (e.g., individual DML or DDL statements). Similar to prior work, what constitutes an input or output of each individual statement is based on the SQL semantics of the statement type (e.g., a \texttt{CREATE TABLE X} statement has \texttt{X} as output); we discuss more on this in~\Cref{ss:lineage-extractor}. 

Note that by using the set union operator to aggregate provenance information, we end up deduplicating multiple instances of an input or output dataset into a single dataset instance (e.g., in our example, if we execute \syncnewsales multiple times, each resulting in executing a query \texttt{INSERT SalesHistory...}, our set union semantics will lead to having \texttt{SalesHistory} as output of \syncnewsales only once). This design enables providing an overview of provenance information with reduced noise (e.g., provenance of a stored procedure shows only single instances of inputs and outputs). At the same time, we still allow drilling down to capture all the details with respect to the alternative instances (e.g., on the provenance of the queries of the stored procedure). We discuss more on this in~\Cref{ss:stitcher}.

Finally, note that although input/output relationships in \atlas can encode what are the input/output datasets of a process, they cannot encode which input contributes to which output. For this, we require a ternary relationship $(i, p, o)$ encoding that input $i$ contributes to output $o$ through process $p$. Since \atlas only supports binary relationships, a common workaround is to introduce this ternary relationship as an attribute on the process (serialized as a dictionary). Importantly, input/output relationships still need to be extracted despite being redundant because other \atlas features (e.g., provenance visualization or label propagation) rely on them. While not ideal, we opt for this workaround in favor of complying with the \atlas open standard.

So far we have discussed about our provenance data model that essentially forms the output of our extraction system. Next, we discuss on query logs that are input to our system. 

\subsection{Query Log} 
\label{s:bg:qlog}

\begin{figure}[t]    
	\centering
	\includegraphics[width=.8\columnwidth]{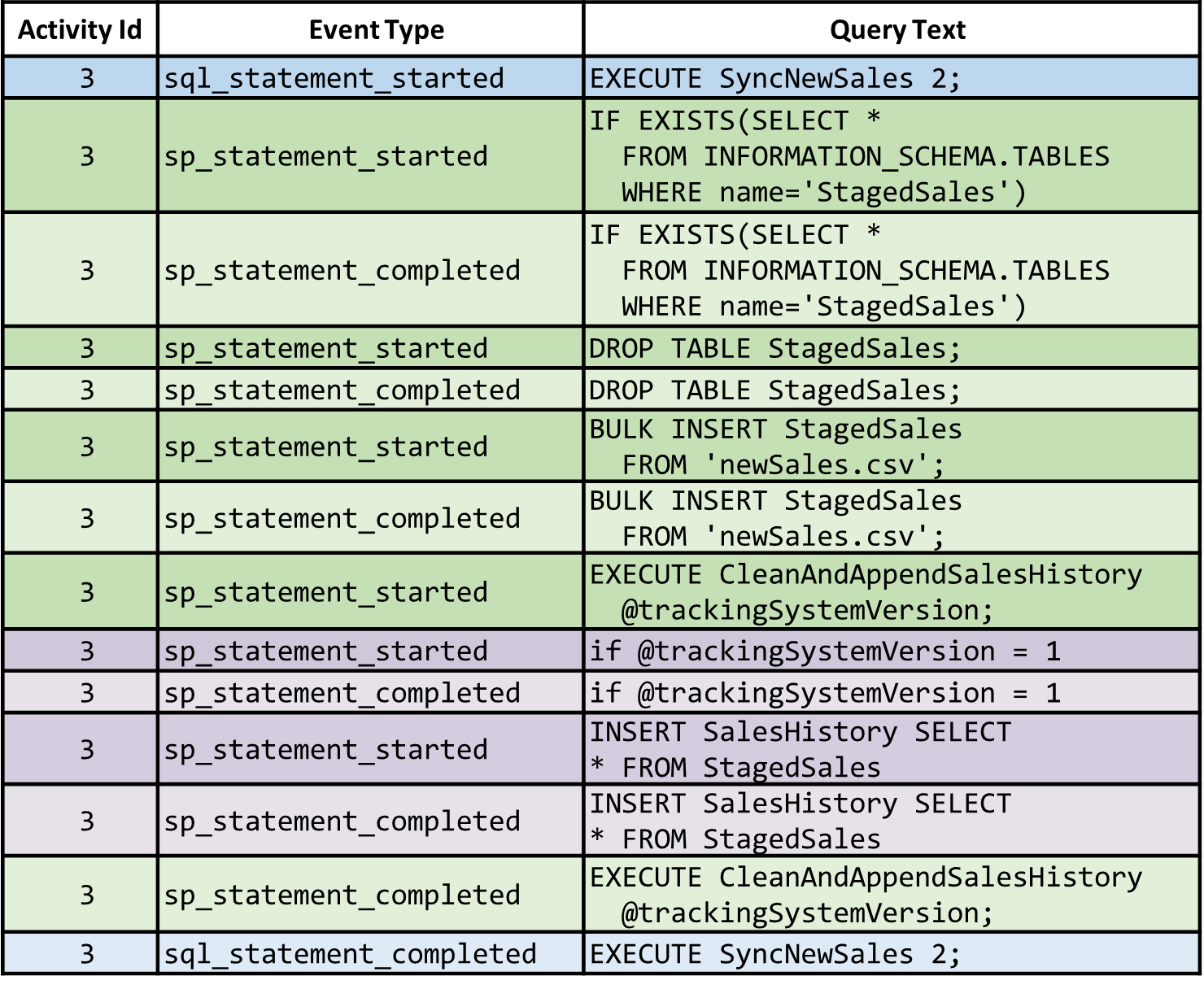}
	\vspace{-1em}
	\caption{\small Events for activity \texttt{EXECUTE \syncnewsales 2} (ordered based on the event triggered time). Each event is associated with activity id, event type, query text, and other metadata that are not shown for brevity (e.g., time triggered, CPU time, or on whose behalf a query was executed).}
	\label{f:ex-log}
	\vspace{-2em}
\end{figure}

We  assume that a database creates a query log as a side effect of query execution. We model the query log as a totally ordered set of events---with ordering based on the time each event was spawned. Next, we discuss on the semantics we require from query logs. To simplify our discussion, \Cref{f:ex-log} shows a query log created by Azure SQL DB for our running example in~\Cref{f:ex-workflow}.

\sstitle{Event types.} Each event is associated with a query run, and can have one of the following two types: 1) \textit{started} and 2) \textit{completed}, indicating the start or completion of the corresponding query run, respectively. Furthermore, we assume that each event is drawn from a collection of event types that encode whether the query is an individual or batch statement. For instance, in our example in~\Cref{f:ex-log}, the entries in the log from Azure SQL DB contain the event types \texttt{sql\_statement\_[started|completed]}, \texttt{sp\_statement\_[started|completed]}, \texttt{sql\_batch\_[started|completed]} to indicate the start and completion of ad hoc statements, statements that are part of a stored procedure, and batch queries, respectively. Note that a direct annotation on whether a query is a batch or part of a stored procedure is not required (if it can be inferred during extraction), but it simplifies the provenance extraction process.

\sstitle{Event schema.} Each start and completion event can contain a rich set of runtime metadata (e.g., who, when, and from where executed a query run and for how much CPU time). As such, each event can be modeled as a record, where each piece of metadata is represented as an attribute. Note that events can have different schemas depending on whether they correspond to start or completion of a query run, and individual statement or batches. For instance, only completed events contain how much CPU time was required by a query run.

\sstitle{Activities.} Entries in the log are clustered into \emph{activities}, with each activity comprising a set of start and completed events triggered for correlated query runs. In general, such clustering is required to identify the set of dependent query runs of each query run, which is a critical step towards addressing our provenance extraction problem. For instance, in~\Cref{f:ex-log}, all events belong to the same activity identified as 3. (This information is provided in the logs generated by Azure SQL DB if causality tracking is enabled~\cite{url:xevents}. Note that not all databases provide such a capability. Yet every database needs to track this information internally for query execution purposes. Hence, we believe our discussion could also inform what databases need to log for efficient provenance extraction purposes.) Going back to our running example, we use this information to deduce what queries have been run as part of \syncnewsales and \cleanandappendsaleshistory; we discuss this in more detail in~\Cref{ss:activity-collector}.

\subsection{Problems of Focus}
\label{s:bg:problem}

Having defined our model and query log, our goal is twofold:

\stitle{Provenance graph extraction:}   First, we aim to extract the provenance graph based on our model in Section~\ref{s:bg:datamodel}. Formally, given a query log, our goal is to: (1) create the provenance graph $P(V,E)$ by extracting edges $E=\{i \xleftrightarrow{Q} o\, |\, i \in I(Q) \subseteq V\, and\, j \in O(Q) \subseteq V\}$, where $Q$ is either a query run $Q_r$ or its static counterpart $Q_s$, and types of inputs $I(Q)$ and outputs $O(Q)$ are relations (encoding relation-level provenance) or even columns (encoding column-level provenance), and (2) associate nodes and edges in the provenance graph (i.e., inputs, outputs, and queries) with metadata from events. We discuss how \sys addresses this problem in~\Cref{s:onep}.

\stitle{Noise reduction:} Second, recall from ~\Cref{s:intro} that a major limitation of existing dynamic extractors is that they treat every query in a query log as equally important often resulting in noisy provenance graphs (\textbf{L4}). Hence, we aim to develop techniques to reduce the noise in the provenance graph $P(V,E)$ during the extraction process. We describe our optimization techniques in~\Cref{s:opts}.

\section{OneProvenance}
\label{s:onep}

\begin{figure}[t]
	\centering
	\includegraphics[width=.9\columnwidth]{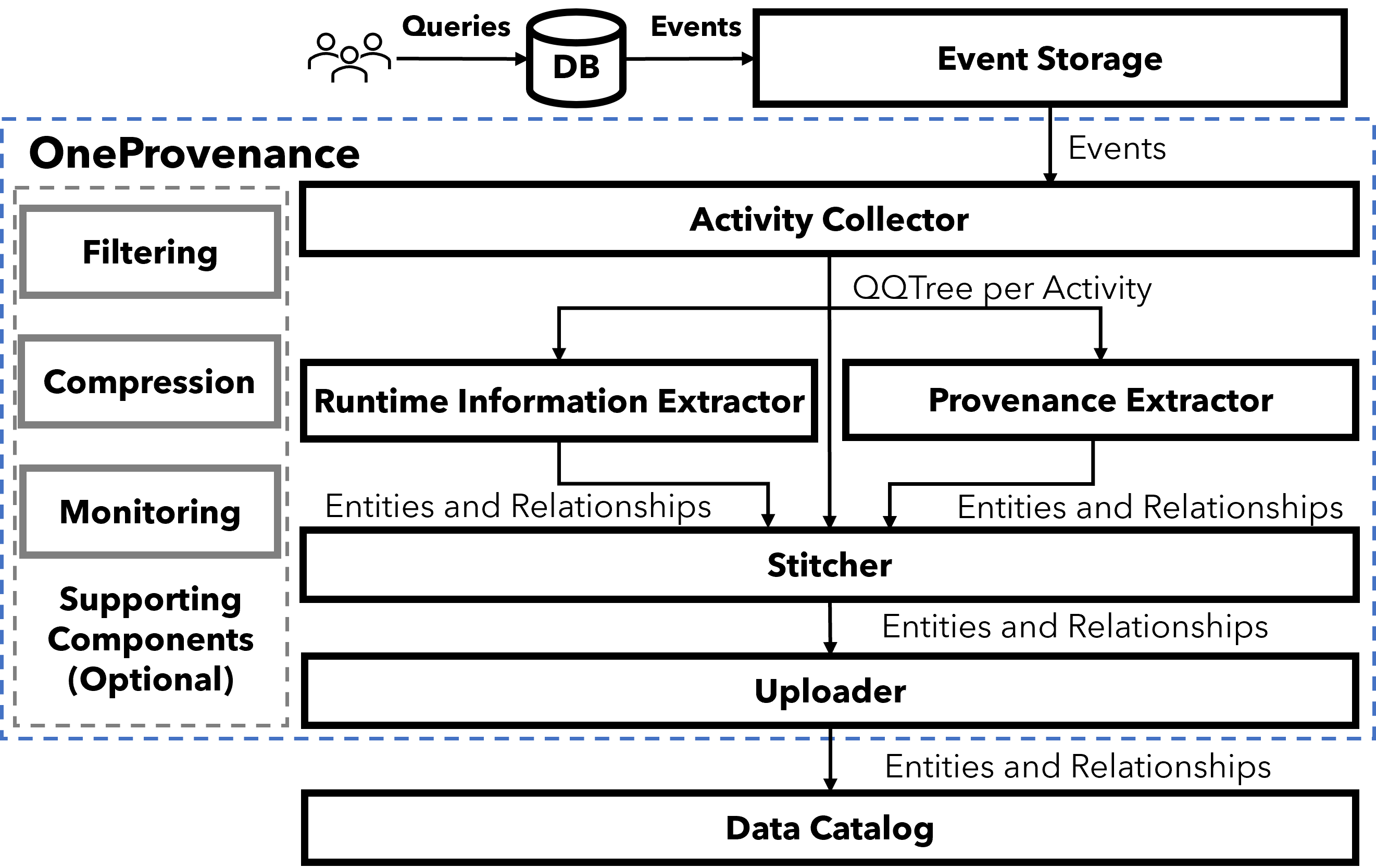}
	\vspace{-1em}
	\caption{\sys Architecture. Main components in \sys's processing flow are in black boxes, and annotated with formats of inputs and outputs. Boxes in grey are optional components, and can be used throughout various components in the main flow.}
	\label{f:sys}
	\vspace{-2em}
\end{figure}

We are now ready to introduce \sys, our provenance extraction engine. Its architecture is shown in~\Cref{f:sys}.

\stitle{Overview.} In short, events emitted by the database due to query execution are stored in an Events Log Storage (\Cref{ss:logstorage}). The \activitycollector of \sys reads these events periodically; identifies SQL activities; and builds an internal representation for each activity, namely, \xruntree, that encodes dependencies between query runs in the activity (\Cref{ss:activity-collector}). For each activity, the identified \xruntree is pushed to the \runtimeinfoextraxtor (\Cref{ss:runtime-info-extractor}) and \lineageextractor(\Cref{ss:lineage-extractor}) components. The former is responsible for extracting queries, query runs, and their relationships based on information encoded in the \xruntree. The latter extracts provenance information from individual statements in the \xruntree. The outputs of both components are encoded according to the \sys data model of \Cref{s:bg:datamodel}. These outputs, along corresponding \xruntree{}s, are then fed to the \stitcher component (\Cref{ss:stitcher}), that stitches the per-statement provenance extracted by the \lineageextractor to the queries and query runs extracted by the \runtimeinfoextraxtor, and aggregates provenance to parent nodes and across runs (per \textbf{L2-3}). The end result is an instance of our provenance data model that gets uploaded to the external Data Catalog through the \uploader component (\Cref{ss:uploader}). 
Finally, we introduce hook points (\Cref{ss:hooks}) into the \sys extraction workflow to provide flexibility and extensibility (per \textbf{L4}).

We next discuss the role of each component and how \sys addresses technical challenges per component. Note that our architecture is agnostic to the database engine since we expect logs with specific semantics (see also~\Cref{s:bg:qlog}). To ease our discussion, however, we consider Azure SQL DB to be the source database, as this is also the one currently supported in production in Purview.

\subsection{Event Logs Storage}
\label{ss:logstorage}

At a high level, queries admitted to a database trigger the generation of events that are persisted by the database in a storage of event logs, as shown in \Cref{f:sys}. We assume that the type of store can be provided as a configuration to the database. For instance, in Azure SQL DB, such events are known as \xevents, and the corresponding \xevent Store can be a local or remote file system, or even a finite memory buffer lying in the database server itself~\cite{url:xevents}. The choice of store is mainly influenced by the application needs and cost constraints. We have opted for Azure Storage~\cite{url:azurestorage} for our Event Logs Storage because this was the cheapest option among alternatives.

\subsection{Activity Collector}
\label{ss:activity-collector}

The \activitycollector is responsible for identifying SQL activities and building \xruntree{}s that encode query execution dependencies in these activities. To do so, it retrieves new events from the Event Logs Storage, parses the events, groups them by the activity they belong to, sorts activities by their trigger time, and constructs a \xruntree per activity based on events of the activity.

To fetch events, the \activitycollector runs periodically (by default every 6 hours). To guarantee that activities that span multiple \sys runs will eventually be processed, all while processing activities only once, \sys uses checkpointing for runs: only events of activities that have at least one event created after the start of the last run of \sys will be processed. Parsing and grouping events is closely tied to the event format. Since our prototype implementation focuses on \xevents, parsing uses XELite~\cite{url:xelite}, and \sys groups activities based on the Activity Id available in each event as discussed in \Cref{s:bg}.

\begin{figure}[t]
	\centering
	\vspace{-1em}
	\includegraphics[width=.85\columnwidth]{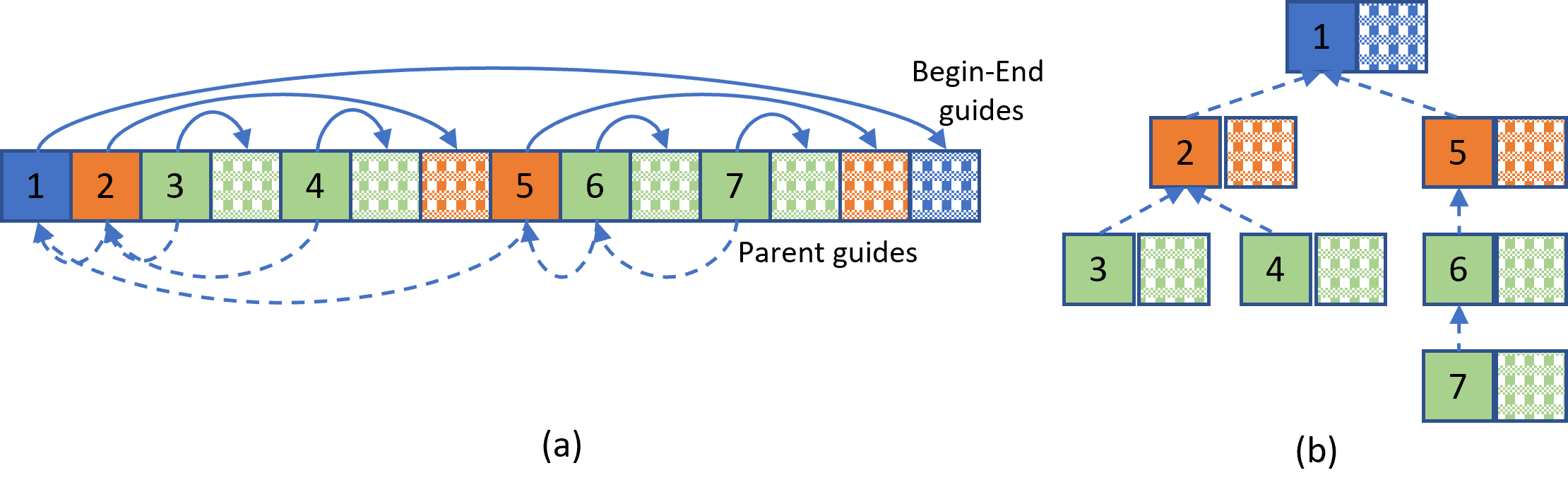}
	\vspace{-1.5em}
	\caption{Example visualization of SQL Activity construction. Solid and patterned fill colors indicate started and completed events, respectively. \textcolor{blue}{Blue}, \textcolor{orange}{orange}, and \textcolor{darkgreen}{green} boxes denote executions of batches, stored procedures or triggers, and individual statements, respectively.}
	\label{f:activity-construction}
	\vspace{-1.5em}
\end{figure}

\Cref{f:activity-construction}a illustrates an event log retrieved by the \activitycollector: each box represents an event, and the color and fill patterns are used to identify event type differences. In particular, \textcolor{blue}{blue}, \textcolor{orange}{orange}, and \textcolor{darkgreen}{green} boxes correspond to events for batches, stored procedures or triggers, and SQL statements, respectively. Execution of each query results in generation of two events: started (solid fill) and completed (patterned fill) events. The \activitycollector uses event attributes to build the \xruntree for each activity, shown in \Cref{f:activity-construction}b.

\begin{figure}[tb!]
	\vspace{-1em}
	\begin{center}
		{\footnotesize
			\begin{minipage}{3.36in}
				\myhrule 
				\vspace{-1ex}
				\mat{0ex}{
					{\bf Algorithm}~\texttt{\xruntree Construction} \\
					\sstab {\sl Input:\/} \= Events $E_A=\{e_1, \ldots, e_n\}$ of a SQL activity $A$ ordered by time \\spawned: $e_1^{time} <\ldots< e_n^{time}$ with $e_i^{time}$ denoting the time $e_i$ was spawned.\\ 
					{\sl Output:} \xruntree $X_A$ for SQL activity $A$. \\
					\bcc \hspace{2ex}\= xStack:Stack[(event, node)] = $\emptyset$\\ 
					\hspace{2ex}\hspace{2ex}\= $X_A$:\xruntree = $\emptyset$ \\ 
					\hspace{2ex}\hspace{2ex}\= curParent:QQTreeNode=nil \\
					\icc\> \ForEach event e \IN $E_A$ \Do \\
					\icc\> \hspace{2ex} \If isStartedEvent(e) \Then \\
					\icc\> \hspace{2ex}\hspace{2ex} node = Node(e, curParent) \\
					\icc\> \hspace{2ex}\hspace{2ex} \If xStack.isEmpty() \Then \\
					\icc\> \hspace{2ex}\hspace{2ex}\hspace{2ex} $X_A$.AddRoot(node)\\
					\icc\> \hspace{2ex}\hspace{2ex}\hspace{2ex} curParent = node\\					
					\icc\> \hspace{2ex}\hspace{2ex} \Else \\
					\icc\> \hspace{2ex}\hspace{2ex}\hspace{2ex} curParent.AddChild(node); \\
					\icc\> \hspace{2ex}\hspace{2ex}\hspace{2ex} \If startsSubTree(e) \Then \\
					\icc\> \hspace{2ex}\hspace{2ex}\hspace{2ex}\hspace{2ex} curParent = node \\
					\icc\> \hspace{2ex}\hspace{2ex} xStack.Push((e, node))\\
					\icc\> \hspace{2ex} \Else \\
					\icc\> \hspace{2ex}\hspace{2ex} (startedEvent, node) = xStack.Pop() \\
					\icc\> \hspace{2ex}\hspace{2ex} \If CheckErrors(startedEvent, node, e) \Then \\ \icc\> \hspace{2ex}\hspace{2ex}\hspace{2ex} Abort() \\
					\icc\> \hspace{2ex}\hspace{2ex} node.SetCompleted(e) \\
					\icc\> \hspace{2ex}\hspace{2ex} \If startsSubTree(startedEvent) \Then \\
					\icc\> \hspace{2ex}\hspace{2ex}\hspace{2ex} curParent = node.parent \\
					\icc\> \If !xStack.Empty() \Then \\ 
					\icc\> \hspace{2ex} Abort() \\
					\icc\> \Return $X_A$ \\
				}				
				\vspace{-5ex}
				\myhrule
			\end{minipage}
		}
	\end{center}
	\vspace{-3ex}
	\caption{\xruntree Construction Algorithm.}
	\label{f:xtreeconstruction}
	\vspace{-2em}
\end{figure}

Algorithmically, a \xruntree is constructed by reading the events of an activity in the order that events were spawned. The corresponding algorithm is shown in~\Cref{f:xtreeconstruction}. When a started event is encountered (lines 4-12), we generate a new node of the \xruntree for this event (line 4), and push the node along the event into the stack (line 12). The new node either becomes the new root of the tree if the stack was empty (lines 5-7), or is appended to the children of the current parent (line 9). Importantly, the new node becomes the current parent (line 11) if the event corresponds to the start of a query that might result in the execution of other queries as part of it (e.g., stored procedure). If a completed event is encountered (lines 14-17), we pop the started event from the stack (i.e., the completed event matches the started event at the top of the stack) and the \xruntree node is initiated. Then, the completed event is stored into the popped node (line 17), and the current parent becomes the parent of the popped node if the latter was starting a subtree (lines 18-19). Finally, note that the algorithm checks for potential errors and aborts accordingly (lines 15-16, 20-21). Errors include log malformation (e.g., log starts with a completed event, or a completed event matches a started event of a different type or query) or started events left in the stack (denoting the activity has not finished executing yet).

\stitle{Time and space complexity.} The \xruntree construction algorithm matches started with completed events to generate \xruntree nodes. It runs in $\mathcal{O}(n)$ time, with $n$ being the total number of events in an activity, by exploiting the ordering of events in the activity (i.e., a completed event matches always the started event at the top of stack). Space-wise, note that only started events are pushed in the stack and generate a node in the \xruntree. Hence, the space complexity is $\mathcal{O}(m)$, with $m$ being the number of started events.

\begin{figure}[t]    
	\centering
	\includegraphics[width=.75\columnwidth]{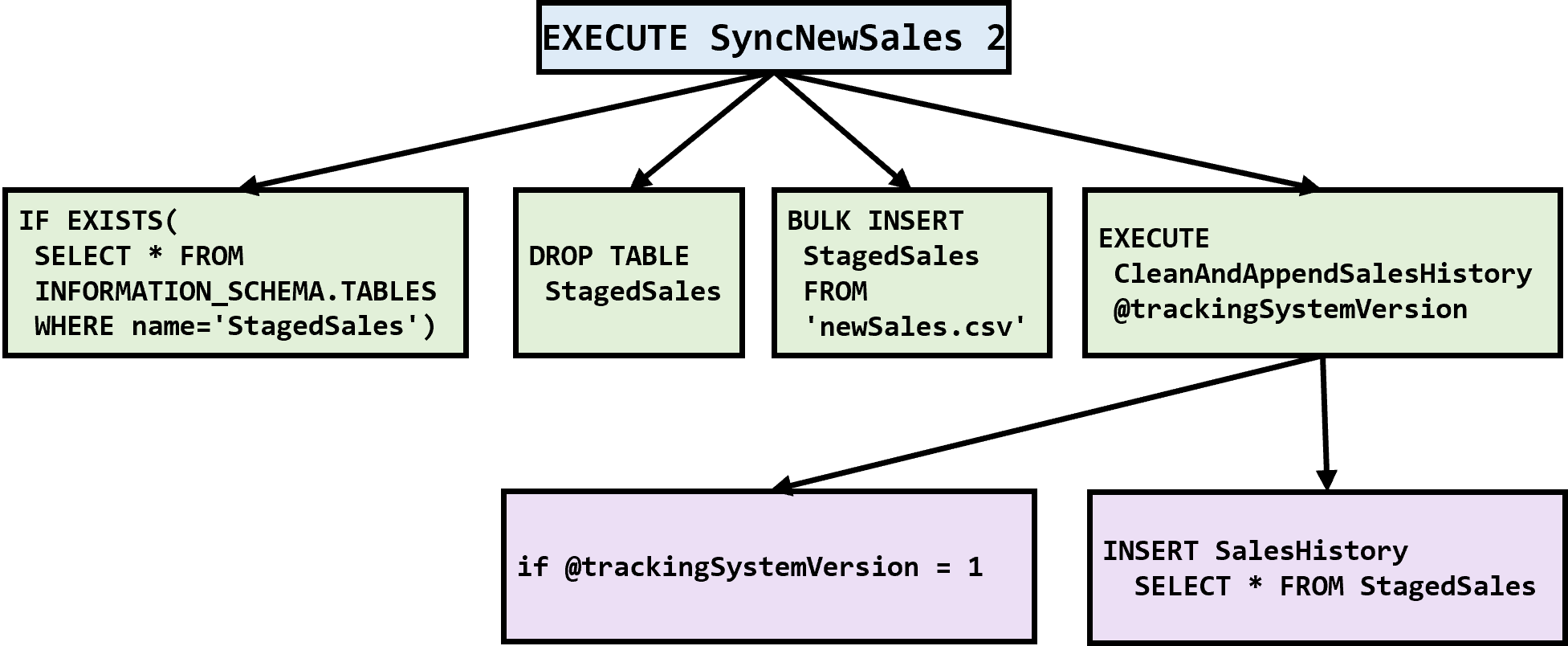}
	\vspace{-1em}
	\caption{\xruntree encoding query dependencies for the activity \texttt{EXECUTE \syncnewsales 2}. In every node, \sys tracks metadata available in the events that the node originated from.}
	\label{f:ex-xtree}
	\vspace{-2em}
\end{figure}

\begin{example} \normalfont\sloppy Consider the execution of \syncnewsales in our running example (\Cref{f:ex-workflow}). In an Azure SQL DB instance, this execution results in the \xevents of~\Cref{f:ex-log}: all statements executed as part of \syncnewsales belong to the same activity. The corresponding \xruntree is shown in~\Cref{f:ex-xtree}. Both log entries and \xruntree nodes have been color-coded to denote matching of started and completed events along what node of the \xruntree they contribute: The parent node denotes the execution of \texttt{EXECUTE \syncnewsales 2}, with children being all queries executed as part of \texttt{EXECUTE \syncnewsales 2} colored in \textcolor{darkgreen}{green}. Since \syncnewsales calls the stored procedure \cleanandappendsaleshistory, queries executed as part of \cleanandappendsaleshistory are included as children of \cleanandappendsaleshistory, and introduce a second nest level in the \xruntree colored in \textcolor{mypurple}{purple}.
\end{example}
\vspace{-1em}

\begin{figure}[t]    
	\centering
	\includegraphics[width=.85\columnwidth]{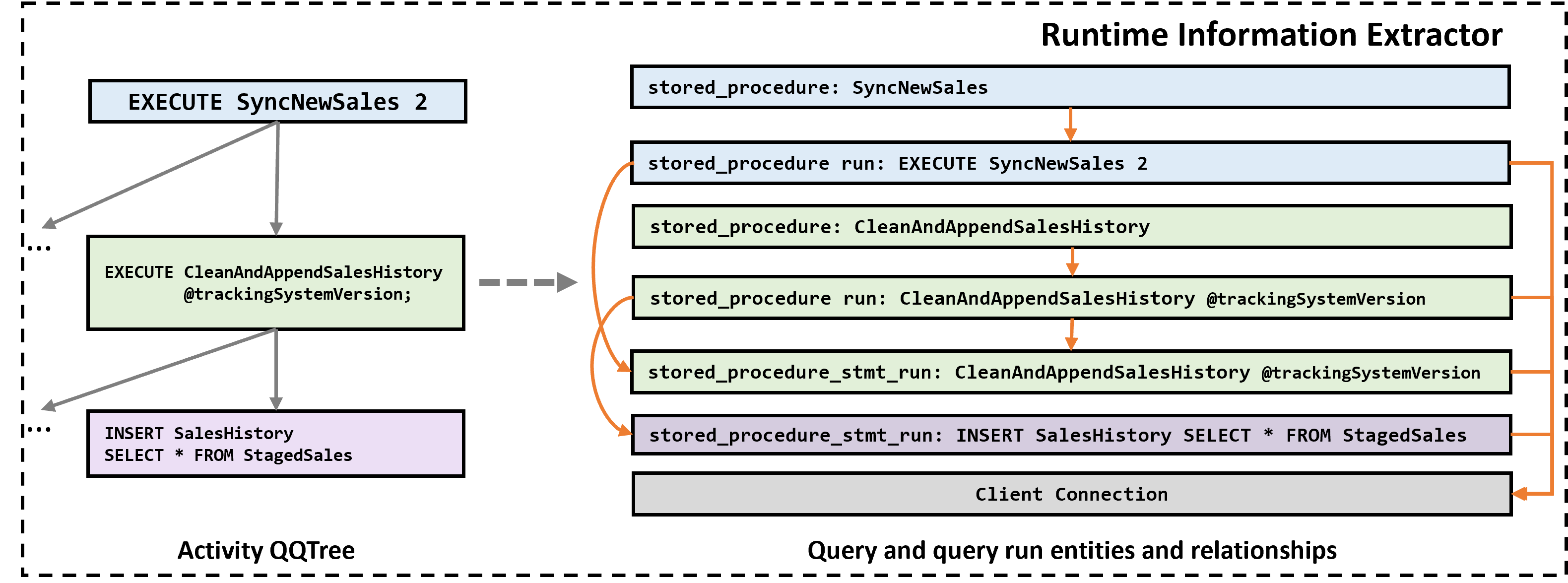}
	\vspace{-1em}
	\caption{The Runtime Information Extractor visits the nodes of \xruntree of each activity to generate entities and relationships related to query and query runs per our provenance model of~\Cref{s:bg:datamodel}.}
	\label{f:rinfo-extractor-ex}
	\vspace{-1em}
\end{figure}

\subsection{Runtime Information Extractor}
\label{ss:runtime-info-extractor}

Given a stream of identified activities, the goal of the \runtimeinfoextraxtor is to instantiate entities and relationships based on \sys's model. To do so, it analyzes events and query dependencies per activity using the corresponding \xruntree. 

More specifically, the extractor visits the nodes of the \xruntree in a DFS traversal. If a visited node corresponds to a stored procedure execution, the extractor generates the corresponding stored procedure and stored procedure run entities (initiating their attributes, such as query text or CPU time, based on event metadata available in the node), along with a relationship between the stored procedure and its runs. Similar would be the case for batches and individual statements. If a statement run is part of a stored procedure (similarly for batch), then a containment relationship is instantiated to encode that the statement is part of a stored procedure run. Finally, note that all run entities are associated with a client connection encoding from what server and application the execution was triggered.

To illustrate this process, consider again our running example.~\Cref{f:rinfo-extractor-ex} shows the entities and relationships generated by visiting (some of) the nodes of the \xruntree of our running example. Visiting the root \texttt{EXECUTE \syncnewsales 2} results in the generation of the stored procedure and stored procedure run entities for \syncnewsales (\textcolor{blue}{blue} boxes) along with a relationship encoding the association between the stored procedure and its run. Similarly, visiting its child generates the nodes associated with \cleanandappendsaleshistory. Note that the statement \texttt{EXECUTE \cleanandappendsaleshistory @trackingSystemVersion} that triggers the stored procedure \cleanandappendsaleshistory is executed as part of \syncnewsales. Hence, besides generating the stored procedure and stored procedure run entities, \sys also generates the stored procedure statement run and the containment relationship with the \syncnewsales run entity. The case for the INSERT statement is similar, resulting in a stored procedure statement run and a containment relationship with its parent stored procedure run \cleanandappendsaleshistory. Finally, all runs are associated with the client connection that initiated the stored procedure execution.

The main focus of the \runtimeinfoextraxtor is on extracting entities and relationships related to processes (queries and query runs). Extracting datasets and provenance relationships is the focus of the \lineageextractor and \stitcher that we discuss next.

\begin{figure}[t]    
	\centering
	\includegraphics[width=.85\columnwidth]{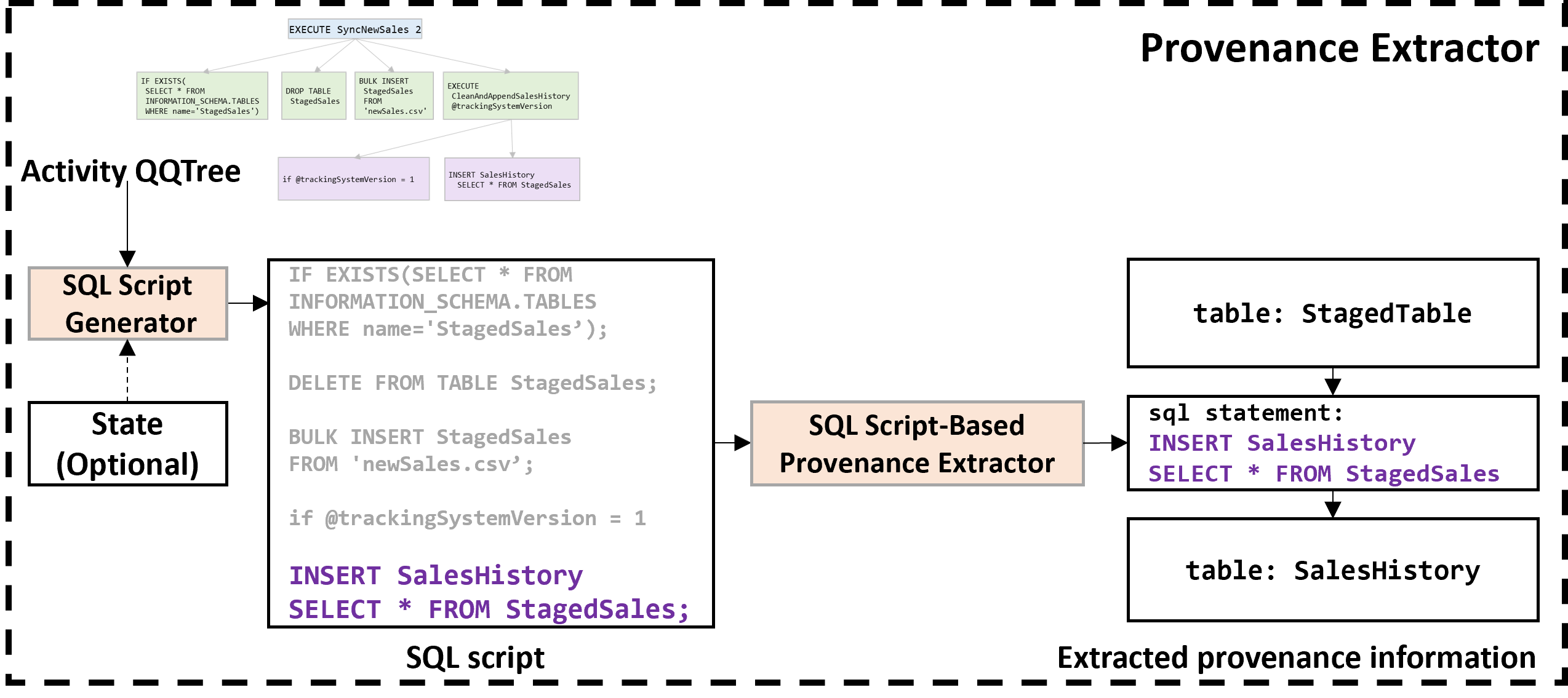}
	\vspace{-1em}
	\caption{The \lineageextractor takes as input a \xruntree and generates a script with the queries of the activity (\sqlscriptgenerator). By analyzing the script it extracts entities (e.g., tables or columns) and their provenance relationships (SQL Script-Based Provenance Extractor). For proper identification of provenance relationships, cataloged information (e.g., view or table definitions) can be embedded in the script by accessing State (i.e., either the database catalog, the catalog populated by previous runs of \sys,  or cached copies of them).}
	\label{f:lineage-extractor-ex}
	\vspace{-2em}
\end{figure}

\subsection{Provenance Extractor}
\label{ss:lineage-extractor}

The \lineageextractor takes the stream of identified activities and extracts provenance relationships and datasets (e.g., tables, views, and columns) according to \sys's data model.

The main components that comprise the \lineageextractor are highlighted in \Cref{f:lineage-extractor-ex}: (1)~SQL Script Generator and (2)~SQL Script-Based Provenance Extractor. The former takes the stream of activities identified by the \activitycollector and generates a SQL script per activity that includes the series of queries executed as part of the activity. The script is generated in a DFS pre-order traversal of the \xruntree that appends the query text of each visited node to the script. Then this script is analyzed statically by the latter component, extracting dataset entities and provenance relationships from it. Importantly, the \sqlscriptgenerator is responsible for location tracking (i.e., mapping queries back to the nodes of the \xruntree they originate from). This is necessary for stitching the extracted provenance information with runtime information extracted by the \runtimeinfoextraxtor, as we will see in~\Cref{ss:stitcher}.

To ensure correctness of the output entities and provenance information, one key technical challenge that provenance extractors need to overcome is to bound objects in a query to catalog objects. For instance, for the query \texttt{INSERT SalesHistory SELECT * FROM StagedSales}, we need to know the schema of \texttt{SalesHistory} and \texttt{StagedSales} to provide column-level provenance. There are many ways bounding information can become available to a provenance extractor; \sys supports the following three cases. 

First, queries appearing in a query log may already be bounded (e.g., * in our example could have been bounded to the actual columns of \texttt{StagedSales}). This would be an ideal situation because it allows a provenance extractor to operate in a stateless fashion. (To be more precise, the ideal situation would be that objects in the query are mapped to their name \emph{and} id in the source database catalog. This is because object names alone may not be unique over time (e.g., if we create, drop, and then create a table with the same name, then the table name may refer to the table before or after the drop).

If bounding information is not available, then \sys needs to infer it. To do so, the \lineageextractor maintains a state that aims to mirror the source database catalog. Hence, a second case supported by \sys is to infer the bounding based on available state. However, it is possible that state is either not available or out-of-sync from the database state (e.g., \lineageextractor runs on logs in a post-mortem fashion when the database catalog is unavailable). Under this third case, \sys extracts provenance with suggestions under ambiguity, resorting to all-inputs-affect-all-outputs in the worst case. 

Overall, we believe that provenance extractors need to support all three scenarios, and customize behavior based on query workloads and log semantics. If a database catalog does not change, then mirroring the catalog in the extractor is a sensible option. Otherwise, either the log provided to the extractor ensures strict serializability~\cite{bernstein:1979:serializability} for catalog replay or queries in the log need to be bounded. If such approaches are deemed costly, provenance extractors should run under best-effort semantics and make suggestions under ambiguity.

\begin{figure}[t]    
	\centering
	\includegraphics[width=.85\columnwidth]{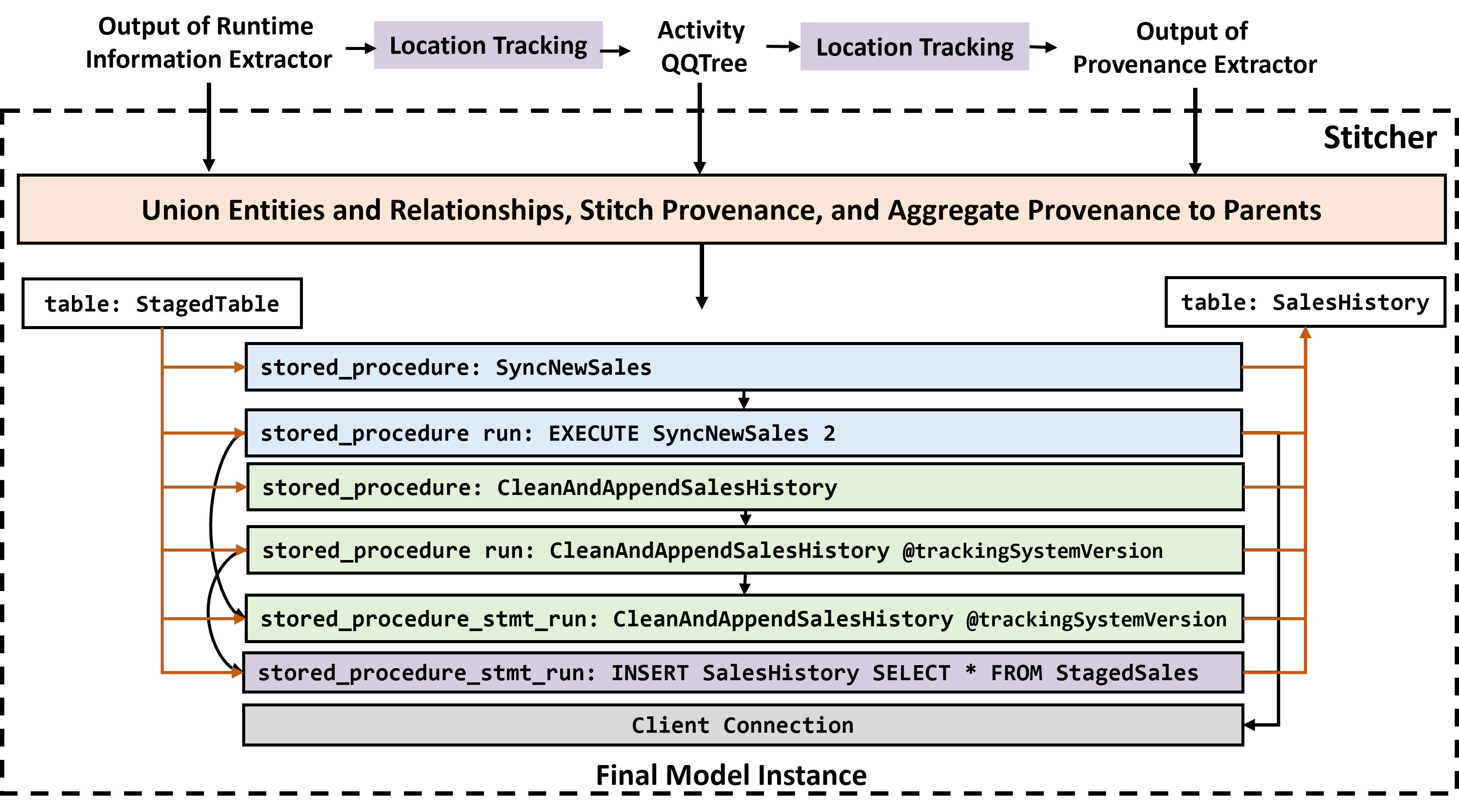}
	\vspace{-1em}
	\caption{\small The \stitcher operates on the outputs of Runtime Information and Provenance extractors: unions the entities and relationships from their outputs, stitches provenance from the \lineageextractor to statements extracted by the \runtimeinfoextraxtor, and aggregates provenance from statements to parent static queries and query runs (e.g., stored procedures and their runs) based on the \xruntree.}
	\label{f:stitcher-ex}
	\vspace{-1.5em}
\end{figure}

\subsection{Stitcher}
\label{ss:stitcher}

The goal of the \stitcher is two-fold: (1)~union the set of entities and relationships identified by the Runtime Information and Provenance Extractors, and (2)~attach and aggregate provenance identified by the \lineageextractor to the process entities identified by the \runtimeinfoextraxtor.

Going back to our running example, consider the outputs produced by the Runtime Information and Provenance extractors that are input to the \stitcher (\Cref{f:stitcher-ex}). For individual statements, the \stitcher attaches the provenance extracted by the \lineageextractor to the corresponding query and query run entities in the output of the \runtimeinfoextraxtor. To do so, the \sqlscriptgenerator performs location tracking to map statements in the SQL script back to the nodes of the \xruntree they came from, as discussed in \Cref{ss:lineage-extractor}. Similarly, the \runtimeinfoextraxtor performs location tracking to map output entities to the nodes they originate from. Hence, the stitching is performed by going from the statements in the SQL script, to the nodes of the \xruntree they originate from, and from there to their corresponding query and query run entities. 

Finally, note that \lineageextractor extracts provenance at the statement level. The \stitcher is also responsible for aggregating the provenance information across runs and to parents by relying on the hierarchical structure of the \xruntree (based on the set union semantics of~\Cref{s:bg:datamodel}). For instance, \stagedsales is aggregated as input to (a) the parent \syncnewsales run from the statement \texttt{INSERT SalesHistory ...}, and (b) the \syncnewsales stored procedure. The end result for our example is shown in~\Cref{f:stitcher-ex}.

\subsection{Uploader}
\label{ss:uploader}

The end-result of the extraction process is the instantiation of a semantically rich data model in the form of entities and related relationships extracted from a series of SQL Activities. The last step of \sys is to upload the instantiated data model to the backend provenance store of a target Data Catalog. Concretely, the main responsibility of the \uploader is to (a) compile our internal data model to the target data model (by default, \sys's output is \atlas-compatible---hence, this step is omitted if the Data Catalog is \atlas-compatible), (b) optimize the upload (i.e., partition entities and relationships for batch uploading), and (c) introduce resiliency (i.e., checkpoints the upload progress and retries failed uploads). Although the \uploader targets Microsoft Purview, \atlas, \openlineage, and \egeria catalogs, it also provides an extensible interface to support other backends.

\begin{figure}[t]
	\centering
	\includegraphics[width=.9\columnwidth]{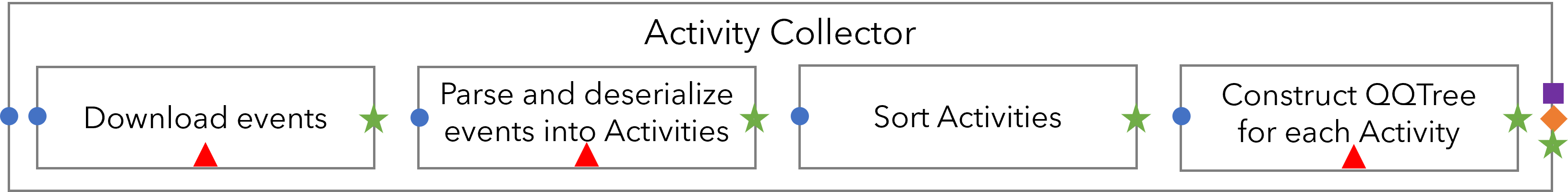}
	\vspace{-1em}
	\caption{Hooks in the logic of the \activitycollector.}
	\label{f:activity-collector-hooks}
	\vspace{-2em}
\end{figure}

\subsection{Hook Points and Code Injection}
\label{ss:hooks}

\sys introduces hook points in its components to enable custom logic injection. These hook points are central both to the extensibility of \sys (to better meet the various needs of end applications and users) and its maintainability. In particular, currently hooks are used extensively for monitoring (e.g., tracking telemetry) and debugging purposes, as well as applying filtering and compression optimizations (discussed in detail in~\Cref{s:opts}). Next, we discuss the hooks in the \activitycollector component; the principles behind hook points of other components are similar.

As shown in~\Cref{f:activity-collector-hooks}, the extraction logic in \activitycollector consists of four sub-components: (1) download batch of events from the Event Logs Storage, (2) parse and deserialize events into activities, (3) sort activities temporally, and (4) construct \xruntree for each activity. The hook points available for injecting external logic are inserted into \activitycollector as well as its sub-components. More specifically, \textcolor{blue}{$\bullet$}s correspond to points at the start of (sub-)components while \textcolor{darkgreen}{$\bigstar$}s are points at the end of (sub-)components. \textbf{\textcolor{violet}{$\blacksquare$}}s correspond to points right before a component is sending its results to other components, while \textbf{\textcolor{orange}{$\blacklozenge$}}s correspond to points after sending these results (i.e., when the control flow returns back to the component). Finally, \textcolor{red}{$\blacktriangle$}s correspond to points right before the end of loops over items of concern in (sub-)components. For instance, we can use the \textcolor{red}{$\blacktriangle$} point when constructing \xruntree{}s to filter \xruntree nodes that are not interesting for end provenance applications.

For each hook point, \sys exposes a programming interface to express injection logic. More specifically, users register functions that are called at the time \sys's execution reaches the hook point. Each point is associated with state that functions can read or write (e.g., the \textcolor{red}{$\blacktriangle$} in the last sub-component of~\Cref{f:activity-collector-hooks} provides read and write access to each \xruntree). Finally, to simplify implementation efforts on hooks, \sys provides iterator-, batch-, and producer-consumer style interfaces for data items in the state of a hook point---hence, enabling familiar relational processing.

\section{Optimizations}
\label{s:opts}

So far, we have presented the workflow of \sys, assuming that every event in the log is equally important. However, based on customer feedback, we found that this assumption is rarely true for real-world provenance applications. This is because events contain noisy data (e.g., information about system maintenance or statistic generation queries) that are not necessarily interesting for business purposes, as well as redundant information (e.g., events emitted by queries that are executed in a loop of a stored procedure). From a provenance extraction perspective, processing all these events would result in additional performance overheads.

In this section, we present simple, yet powerful application-aware optimizations for reducing the emitted noise. These optimizations have yielded up to four orders of magnitude performance improvements on provenance extraction from customer production workloads (similar to our experiments). Provenance applications of such customers include end-of-fiscal-year audits, data observability (e.g., are outputs of a stored procedure updated?), data estate understanding (e.g., what datasets are produced by a stored procedure?), context-based analytics~\cite{DBLP:conf/cidr/HellersteinSGSA17}, and impact/root cause analysis for debugging.

As shown in~\Cref{f:sys}, the main items in the flow of \sys are events, queries, activities (each modeled as a \xruntree or series of events), client connections, and entities and relationships. We design a set of, highly-effective, filters that can be applied on such items throughout the flow of \sys:

\stitle{Loop compression.} Stored procedures that execute queries iteratively in a loop introduce redundancies in the extracted provenance information. Such loops can overwhelm a provenance extraction system, all while not capturing any new provenance information. To prevent this issue, \sys introduces a compressor that groups together nodes in the \xruntree originating from multiple iterations of the same loop. In particular, \sys can be configured to keep only the latest $k$ iterations of the loop. By leveraging this compressor, we avoid calling the \lineageextractor with duplicate queries, and thus, overloading the data catalog with redundant information. Additionally, this compressor reduces the overall working memory footprint per activity of \sys.

\stitle{Drop uninteresting queries.} \sys can determine whether to capture the provenance of a query based on whether the query is interesting or not, which is typically driven by user interests and applications needs. In particular, queries can be matched based on several factors, including their type (e.g., \texttt{SELECT} or CRUD statements), syntax tree (e.g., queries to compute statistics in Azure SQL DB are structured as \texttt{SELECT STATMAN(...) FROM T}), whether they access tables and columns (e.g., \texttt{SET @a=2}), or the query text itself (e.g., queries that do not match a regular expression). For matching queries, we can either (a)~drop the \xruntree node corresponding to the query altogether or (b)~consider the node in \runtimeinfoextraxtor and ignore it in \lineageextractor (e.g., \texttt{SET @a=2} can be filtered out from the \lineageextractor since it is not interesting from a provenance perspective, but still be pushed to \runtimeinfoextraxtor for metadata extraction).

\stitle{Drop uninteresting activities.} \sys can also drop activities if they contain uninteresting queries. For instance, filtering out activities that do not contain DDL queries or stored procedure executions is common in \sys deployments.

\stitle{Filters on event metadata.} \sys supports expressing conditions on event metadata to filter out events or activities. For instance, a condition of the form \texttt{client\_app\_name = 'SSMS' or username = 'sa'} can filter out events for queries executed from the \texttt{SSMS} application or by the system administrator \texttt{sa}. Furthermore, recall that activities can be modeled as an ordered set of events (\Cref{s:bg}). As such, conditions on event metadata can be used to drop activities as opposed to individual events (e.g., if we want to focus only on activities with long running queries, we can filter out an activity if \texttt{duration} $<$ X seconds for all completed events).

\stitle{Filters on activities from uninteresting connections.} Based on the above filters on event metadata and activities, we can construct more sophisticated filters. 
In particular, \sys can be configured to filter out an activity if it is coming from an uninteresting connection. A connection is considered uninteresting if (a)~it has no interesting queries or 
(b)~has interesting queries but it does not include any of their last $K$ executions (e.g., last $K$ executions of interesting stored procedure $X$). Note that this optimization guarantees extraction from at least the last $K$ executions of an interesting query but \sys can still extract provenance from more than $K$ executions. This is because another interesting query (e.g., stored procedure $Y$) may be in the connection, rendering the connection interesting (e.g., the execution of $Y$ is in the last $K$ executions of $Y$). Under such a case, \sys can be configured to either consider all queries interesting or keep only the ones matching the last $K$ semantics (e.g., keep $Y$ and filter out $X$). The former is important to provide execution context (e.g., stored procedure $Y$ was called after executing $X$) while the latter is preferable for further noise reduction when no such context is necessary. As we will see in the experiments, the overall effect is a substantial reduction in noise and a significant improvement in the end-to-end extraction latency.

\stitle{Drop levels of aggregation.} Different catalogs and provenance applications may require a different subset of entities and relationships generated by \sys. For instance, applications may not be interested in provenance at the SQL statement level but rather at coarser levels (e.g., stored procedure level), and vice versa. To address such scenarios, \sys can be configured to emit provenance at different levels of aggregation. 

\stitle{Drop events.} So far, we have assumed that logs emitted by the database system are complete, and that filters on events preserve the semantics of ~\Cref{s:bg:qlog}. However, for certain workloads (e.g., heavy transactional), database engines will emit a significant amount of events. Even though \sys optimizes the extraction process by relying on logs with query text (as opposed to query plans), such workloads can still introduce high overhead to both the database and provenance extraction engines. Under such loads, \sys can continue operating by dropping events in favor of avoiding adding overheads on the database. In our experiments, we will show how event retention options supported by \xevents (dropping events based on event buffer availability) can allow \sys to process high (transactional) loads.

\section{Integration with Purview}
\label{s:integration}

Microsoft Purview is a governance platform offering by Microsoft, that allows organizations to govern (e.g., catalog, overview, secure, analyze, and audit) their data estate. To extract metadata and provenance, Purview provides a rich collection of extractors. Each such extractor connects to an underlying data system to extract metadata and provenance. Extraction can be scheduled either as one-off or recurring. The output of extractors are metadata and provenance modeled based on Apache Atlas-based data models. Finally, Purview ingests the instantiated data models and stores the extracted metadata and provenance in its underlying Data Catalog, on top of which it exposes data governance functionalities.

Based on this design, the integration with \sys is straightforward. Recall \sys currently supports extraction from Azure SQL databases. When a customer requires dynamic provenance, Purview sets up an \xevent session in the corresponding Azure SQL database, and the database starts emitting \xevents in Azure Storage. (Note that for security purposes the blob storage is owned by customers and managed by Purview). Then, Purview schedules \sys to run periodically (currently, every 6 hours). When executed, \sys analyzes the underlying logs, as discussed in~\Cref{s:onep}, and pushes the extracted metadata and provenance to Purview. Regarding optimizations, Purview employs all optimizations discussed in~\Cref{s:opts} to decrease the noise of the extraction as much as possible.

In particular, currently Purview enables loop compression with the strictest setting (i.e., keeping only the latest iteration). Furthermore, Purview filters activities based on a set of more than 40 uninteresting queries and query templates including ones for statistics collection, system administration tasks, or connection identification. Moreover, it keeps only the activity with the latest execution of a process (e.g., latest stored procedure execution); all previous activities with the same interesting process are filtered out (i.e., operates under the semantics for as much as possible noise reduction discussed in ``Filters on activities from uninteresting connections'' in ~\Cref{s:opts}). Also, Purview does not enable dropping events by default to avoid missing important provenance information. This setting is used by default for customers that have heavy transactional workloads upon request. Finally, note that users can also alter filters, as we also perform in our experiments.

\section{Experiments}
\label{s:exps}

\newcommand{\sessionreliable}{\texttt{NO\_EVENT\_LOSS}\xspace}
\newcommand{\sessionunbound}{\texttt{ALLOW\_SINGLE\_EVENT\_LOSS}\xspace}
\newcommand{\xequeryplan}{\texttt{query\_post\_execution\_plan\_profile}\xspace}
\newcommand{\sessionqplan}{\texttt{QPlan}\xspace}
\newcommand{\sessionqtext}{\texttt{QText}\xspace}

\newcommand{\sessionplan}{\texttt{Plan-R}\xspace}
\newcommand{\sessionplanel}{\texttt{Plan-LA}\xspace}
\newcommand{\sessiontext}{\texttt{Text-R}\xspace}
\newcommand{\sessiontextel}{\texttt{Text-LA}\xspace}
\newcommand{\sessionfnel}{\texttt{Text-FN}\xspace}
\newcommand{\todo}{\textcolor{red}{TODO}\xspace}
\newcommand{\tpm}{TPM\xspace}
\newcommand{\azsqldb}{Azure SQL DB\xspace}
\newcommand{\procbench}{SQL-ProcBench\xspace}
\newcommand{\tpcc}{TPC-C\xspace}
\newcommand{\tpch}{TPC-H\xspace}
\newcommand{\tpcds}{TPC-DS\xspace}
\newcommand{\sproc}{\texttt{SP}}
\newcommand{\lgread}{\texttt{LgRead}\xspace}
\newcommand{\lgpars}{\texttt{LgPars}\xspace}
\newcommand{\xrunt}{\texttt{QQT}\xspace}
\newcommand{\provex}{\texttt{ProvEx}\xspace}
\newcommand{\hammerdb}{HammerDB\xspace}

\newcommand{\sysp}{OneP\xspace}
\newcommand{\qplan}{QPlan\xspace}

We now present our thorough evaluation of \sys with the goal to (a) compare \sys with state-of-the-art extraction techniques and (b) demonstrate the benefits of our optimizations on performance improvement and noise reduction. 

We begin by briefly describing our experimental setup.

\stitle{Workloads and databases.} For our experiments, we generate workloads using (1)~\procbench~\cite{procbench2021}, (2)~\tpch, (3)~\tpcds and (4)~\tpcc benchmarks. These workloads provide a mix of both real-world and realistic OLAP and OLTP use cases, to help us show the application of \sys and its performance across a wide spectrum of database workloads. (Note that Microsoft Purview does not log customer workloads internally---to comply with privacy requirements of customers. Moreover, the extracted provenance information is only accessible by authorized customers. Hence, it is infeasible to run experiments using real customer workloads. Through the workloads of our experiments, however, we have reproduced the key insights that we observed in production workloads, drove the design of \sys, and discussed throughout the paper.)

\procbench is designed using insights derived from an analysis of SQL queries, UDFs, triggers and stored procedures in 6500 real world applications~\cite{procbench2021}.
The workload uses the \tpcds dataset and comprises 63 stored procedures, out of which we selected 35 that can be run multiple times. In our experiments, we generate the \tpcds database with scale factor set to 1. For a database of this size, we observed an average of 207 SQL statement runs for the selected stored procedures, 
including (1)~8 statements originating from nested triggers, (2)~48 statements originating from UDFs, (3)~33 total loop iterations, and (4)~up to 6 levels of nested dependencies.

\tpch and \tpcds are standard benchmarks for performance evaluation of decision support systems. They consist of 22 and 99 ad hoc analytical queries, respectively. For workloads using these queries, we generated the corresponding TPC-H and TPC-DS databases with scale factors set to 1 and 10. (Insights on \tpch and \tpcds are similar for both scale factors. As such, we report results mainly on TPC-H with scale factor 1 to avoid redundant insights.)

\tpcc is a standard benchmark for performance evaluation of OLTP systems. In contrast to prior analytical workloads, that are the traditional focus of dynamic provenance extraction systems, the transactional load of \tpcc serves as a stress-test that can help us identify the extent of our coverage over high-load workloads. It involves a mix of 5 concurrent transactions of different types and complexity. We generated a database with 40 warehouses, and observed 109 statement runs per transaction on average. The workload mainly comprises many low-latency simple statements, but also includes \texttt{IF} conditions and up to 16 iterations of \texttt{WHILE} loops .

\stitle{Workload Generator.} We used \hammerdb~\cite{url:hammerdb:website,url:hammerdb:github}, an open source benchmarking tool hosted by TPC, to generate and run workloads based on \tpcc, \tpch, \tpcds, and \procbench. For workloads from \tpcc and \tpch, \hammerdb builds a configurable \#client threads to concurrently run a configurable \#transactions (for TPC-C) or \#queries (for TPC-H). Also, we extended \hammerdb to generate workloads from \procbench and \tpcds, with same configurations (i.e., \#client threads, \#transactions, \#queries). 

\stitle{Platform.} All workloads are run against a serverless \azsqldb instance~\cite{url:sqlmi}, with 8 cores and 24 GB memory, which emits logs to an Azure Storage account. \hammerdb and \sys are installed on a Standard D8s v3 Azure VM (8 cores, 32 GB memory). All resources and services are deployed in the same Azure region.

\stitle{Outline.} We start our discussion by breaking down the performance of \sys in comparison with state-of-the-art prior work (\Cref{s:exp-sys-perf}), followed by experiments highlighting the benefits of our optimizations (\Cref{s:exp-opts}), and concluding with a discussion comparing \sys with SAC~\cite{tang:2019:sac}, Spline~\cite{scherbaum2018spline}, and the OpenLineage Spark extractor~\cite{url:openlineage} (\Cref{ss:exp:incomparison}). We provide detailed settings and compared techniques inline per experiment.

\subsection{\sys Performance Breakdown} 
\label{s:exp-sys-perf}

We start our experiments by a performance breakdown on the different components of extraction.

Note that our goal is to compare \sys against state-of-the-art techniques for the provenance capture problem of our focus. Such techniques are employed by dynamic provenance extraction systems (e.g., SAC~\cite{tang:2019:sac}, Spline~\cite{scherbaum2018spline}, and OpenLineage for Spark~\cite{url:openlineage}). These systems, however, target Spark as their source of query event logs. In contrast, \sys targets \azsqldb. Differences in logging by Spark and \azsqldb, and overall network topology of systems (e.g., SAC and Spline run in the master node of Spark, OpenLineage uses Azure functions, and \sys runs outside of \azsqldb) are the main reasons why performance comparisons do not reveal meaningful insights on the core provenance capture problem of our focus.

At their core, however, these systems operate on query plans. As such, we alter \sys to operate on plans to mimic the behavior of these extractors in \azsqldb, and perform meaningful comparisons. We denote this system (\sys with physical plans as input) as \texttt{QPlan} and compare it with \sys. (A discussion on SAC, Spline, and OpenLineage is included in~\Cref{ss:exp:incomparison}.) To better understand overheads on source databases, we also compare \sys and \texttt{QPlan} with query execution without provenance capture (denoted as \texttt{Baseline}).

\subsubsection{Logging} 
\label{ss:exp-xe-session}

The first component where \sys differs from prior work is on logging events. Prior work uses query plans whereas \sys exploits only query text. Hence, next, we aim to compare the overheads on database systems among logging query plans (prior work) and query text (\sys).

\begin{figure}[t]
	\centering
	\begin{subfigure}[b]{0.3\columnwidth}
		\centering
		\includegraphics[width=\columnwidth]{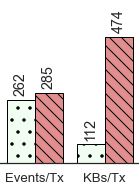}
		\caption{\small \tpcc}
		\label{f:tpcc-tx-xe-storage}
	\end{subfigure}
	\hfill
	\begin{subfigure}[b]{0.3\columnwidth}
		\centering
		\includegraphics[width=\columnwidth]{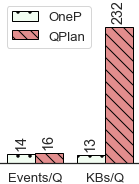}
		\caption{\small \tpch}
		\label{f:tpch-xe-storage-8}
	\end{subfigure}
	\hfill
	\begin{subfigure}[b]{0.3\columnwidth}
		\centering
		\includegraphics[width=\columnwidth]{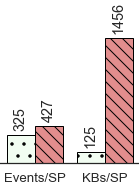}
		\caption{\small \procbench}
		\label{f:profile-pb}
	\end{subfigure}
	\caption{\small Characteristics of query text (\sysp) vs. query plan (\sessionqplan) logging across different workloads. Characteristics include number of events and required storage (in KBs) per transaction, query, or stored procedure. The results show that \sessionqplan events are significantly larger.}
	\label{f:xe-storage}
	\vspace{-1em}
\end{figure}

\stitle{In comparison.} \sloppy More specifically, recall from~\Cref{s:bg:qlog} that \azsqldb provides a configurable query log architecture that allows us to log events of different types. \sys logs \texttt{started} and \texttt{completed} events for executions of \texttt{sql\_statement}, \texttt{sp\_statement}, and \texttt{sql\_batch}. To mimic the behavior of state-of-the-art dynamic provenance extractors that rely their extraction on plans, we introduce \qplan that logs the \xequeryplan event type that carries query plans. Note that \qplan logs the \xequeryplan event type on top of the rest of the event types because the query text and runtime metadata (that are carried only in the rest event types) need to be in the output provenance models (of both ours and prior work) anyways. To compute the overhead of each technique, we use the performance of the database with logging turned off as \texttt{Baseline}.

\stitle{Metrics.} We compare \sys and \qplan on logging based on their I/O requirements and database overheads. For the former, we report the size of logs (in KBs) and number of events per granularity of interest (i.e., KBs and \#Events per transaction for TPC-C, query for TPC-H, and stored procedure for \procbench). For the latter, we rely on transaction rate (\tpm) metric,  average execution latency, and database CPU utilization.

We start our comparison on I/O requirements for the different workloads of our experiments (\Cref{f:xe-storage}).

\stitle{I/O requirements.} Across workloads we observe that \qplan logs a few more events per granularity of interest. For instance, \tpcc transactions run 109 SQL statements on average---hence, the database emits at least 218 started, completed events. \qplan captures 263 events due to the extra logging of \xequeryplan. As query plans are intrinsically large, \sessionqplan logs require up to \textasciitilde18X larger storage compared to \sys logs. This means query plan logging exhibits a very high demand for critical event buffer and I/O resources, that put prohibitive pressure on the database engine and event storage (esp. for high-load workloads). (Note that allocating a large event buffer is not recommended~\cite{url:createeventsession}, and, indeed, in our experiments increasing the buffer did not yield better results for plan-based logging.) 

\begin{figure}[t]
	\centering
	\begin{subfigure}[b]{0.3\columnwidth}
		\centering
		\includegraphics[width=\textwidth]{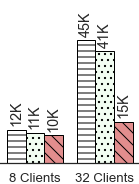}
		\caption{\small Avg. TPM}
		\label{f:xe-impact-tpm-8}
	\end{subfigure}
	\hfill
	\begin{subfigure}[b]{0.3\columnwidth}
		\centering
		\includegraphics[width=\textwidth]{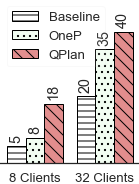}
		\caption{\small Avg. DB CPU load}
		\label{f:xe-impact-cpu-8}
	\end{subfigure}
	\hfill
	\begin{subfigure}[b]{0.3\columnwidth}
		\centering
		\includegraphics[width=\textwidth]{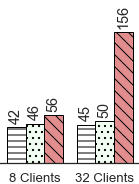}
		\caption{\small Avg. Tx Latency}
		\label{f:xe-impact-latency-8}
	\end{subfigure}
	\caption{\small Comparison of transactions/min (TPM), CPU load, and Tx latency overheads of query text (OneP) vs. query plan (\qplan) logging over no logging (Baseline) for 8 and 32 clients running \tpcc workloads.}
	\label{f:xe-impact-8mb}
	\vspace{-2em}
\end{figure}

Next, we analyze the overheads of logging on database performance (\Cref{f:xe-impact-8mb}) while increasing the number of client threads to collect additional data points. For this experiment, we focus only on \tpcc: due to the high-load and low-latency requirements of transactional workloads we consider \tpcc a stress test in our setup.

\stitle{Logging's impact on query performance.} \Crefrange{f:xe-impact-tpm-8}{f:xe-impact-latency-8} show that logging query plans (\qplan) leads to significantly lower \tpm (up to 3$\times$ lower),
slower transaction execution (up to 3$\times$ increase in transaction execution latency), and a very high CPU utilization (up to 2$\times$ increase).  
Finally, \Crefrange{f:xe-impact-tpm-8}{f:xe-impact-latency-8} highlight that the performance metrics of \sys are comparable to those of the baseline and that the database is able to capture transaction events without paying the reliability penalty as much.

\takeaways{Our results highlight that relying on query text improves significantly the logging performance, database overhead, and overall costs required by provenance extraction systems in comparison to relying on plans (up to 18$\times$ improvements on storage). Our conclusion is that provenance extraction systems relying on query text are low cost solutions and more practical overall.}

\subsubsection{\sys Components} 
\label{ss:exp-sys-perf}

We now focus on the performance evaluation of components of \sys (both in isolation and end-to-end). Our analysis mainly focuses on understanding (1) the performance of \sys and its components (2) improvements of \sys over processing query plans for provenance extraction of prior work (through \qplan) and (3) the performance of aggregating provenance information.

\stitle{Benchmarks.} For this analysis, we used \hammerdb to generate and run 25 iterations of \tpch and \procbench query sets. These workloads can better reveal the performance of \sys components and the overheads of \qplan, since they contain complex analytical queries. Furthermore, we run the \tpcc workload with \#transactions ranging between 8K and 32K. With this workload we aim to test how \sys scales as query load increases.

\stitle{In comparison.} In this experiment, we run \sys with two optimizations on. More specifically, we enable the filtering out of activities from uninteresting connections and loop compression optimizations. For both, we enable their most aggressive filtering out options (i.e., admit only activities with latest runs of \sproc{}s and last loop iterations). These optimizations and their configurations are on by default in  production deployments and, as such, better reflect \sys's performance. (We discuss on the performance of \sys with and without these optimizations in~\Cref{s:exp-opts}.) Finally, we compare \sys with \qplan. Note that, to gain meaningful insights on overheads of processing query plans, we enable the same set of optimizations for \qplan.

\stitle{Components.} We breakdown the latency of \sys to sub-components of the Activity Collector: (1) \lgread (download query logs from event storage), (2) \lgpars (deserialize events in the downloaded logs), (3) \xrunt (analyze events and build \xruntree{}s), and rest \sys components: (4) \provex (Provenance Extractor), (5) Stitcher, and (6) RInfo (Runtime Information Extractor).

\begin{figure}[t]
	\centering
	\begin{subfigure}[b]{\columnwidth}
		\centering
		\includegraphics[width=\columnwidth]{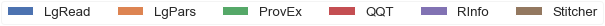}
		\label{f:e2e-legend}
		\vspace{-1.5em}
	\end{subfigure}
	\begin{subfigure}[b]{0.32\columnwidth}
		\centering
		\includegraphics[width=\columnwidth, valign=t]{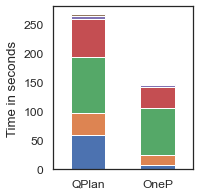}
		\vspace{.4em}
		\caption{\small \procbench}
		\label{f:e2e-pb}
	\end{subfigure}
	\hfill
	\begin{subfigure}[b]{0.32\columnwidth}
		\centering
		\includegraphics[width=\columnwidth, valign=t]{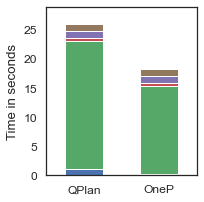}
		\vspace{.4em}
		\caption{\small \tpch}
		\label{f:e2e-tpch}
	\end{subfigure}
	\hfill
	\begin{subfigure}[b]{0.325\columnwidth}
		\centering
		\includegraphics[width=\columnwidth, valign=t]{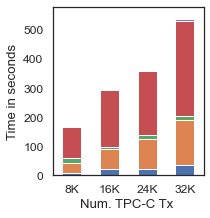}
		\caption{\small \tpcc}
		\label{f:e2e-tpcc}
	\end{subfigure}
	\vspace{-1em}
	\caption{\small Latency (end-to-end and by component; measured in seconds) of \sys and \qplan on (a) \procbench and (b) \tpch, and \sys latency while increasing \#transactions in (c) \tpcc. Main results include: 1)~processing query plans (\qplan) is costlier than query text (\sys), 2)~latency of aggregating provenance is negligible compared to other components, and 3)~latency of \sys increases linearly with query load.}
	\vspace{-1em}
\end{figure}

\stitle{Query plan processing overheads.} As seen in~\Cref{f:e2e-pb}, processing plans (\qplan) is $2\times$ slower compared to \sys's processing query text.
This is because the tasks \lgread, \lgpars, and \xrunt{} for \qplan have to process 10$\times$ additional log bytes and end up dominating the end-to-end latency.
Extraction from \tpch logs witnesses the same overheads, shown in~\Cref{f:e2e-tpch}.
Recall that the \tpch query logs are rather small (e.g., logs corresponding to 100 iterations of the 22 queries are less than 30 MB). Consequently, \lgread, \lgpars, and \xrunt{} (\Cref{ss:lineage-extractor}) terminate rather quickly. Finally, for both workloads \provex has sizeable latency requirements; this is expected since \tpch and \procbench contain complex analytical queries. Interestingly, we observed that the latency of \provex for \qplan is a bit higher than the one of \sys primarily due to the difference in the sizes of plans and query texts.

\stitle{Provenance aggregation.} As also seen in~\Cref{f:e2e-pb}, the latency required for aggregating provenance information (Stitcher) is negligible in comparison to the ones required by other components. As such, \sys addresses the limitations \textbf{L2-3} of prior work originating from lack of aggregations (as discussed in~\Cref{s:intro}) by incurring a negligible overhead in provenance extraction. 

\stitle{Stress test-Latency under query load increase.} \Cref{f:e2e-tpcc} illustrates that \sys remains stable when subjected to higher loads, and the end-to-end latency increases linearly with the load. (\qplan incurs prohibitive costs for \tpcc as we discussed in~\Cref{ss:exp-xe-session} and we omit its performance). Furthermore, an interesting insight from this experiment is that the extraction engine spends the majority of its time in \lgpars{} and \xrunt{} tasks. This is because \tpcc comprise of many low latency queries which are executed repeatedly, thousands of times. In such a workload, our default optimizations manage to filter out repetitive queries. These optimizations are applied right after \xrunt (i.e., after the Activity Collector)---leading to reduced load for \provex and Rest and explaining why \lgpars{} and \xrunt{} dominate the latency.

\takeaways{Overall, our results highlight (1) the importance of avoiding query plans for provenance extraction, (2) that aggregating provenance has negligible overhead, and (3) \sys is an efficient provenance extraction system to the extent that it can support even high-load transactional workloads.}

\subsection{Optimizations} 
\label{s:exp-opts}

\begin{figure}[t]
	\centering
	\begin{subfigure}[b]{0.41\columnwidth}
		\centering
		\includegraphics[width=\columnwidth]{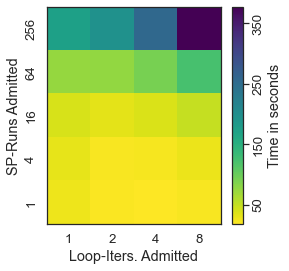}
		\caption{\small Total Execution Time}
		\label{f:topk-heatmap-time}
	\end{subfigure}
	\hfill
	\begin{subfigure}[b]{0.41\columnwidth}
		\centering
		\includegraphics[width=\columnwidth]{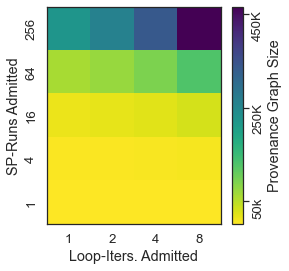}
		\caption{\small Provenance Graph Size}
		\label{f:topk-heatmap-size}
	\end{subfigure}
	\vspace{-1em}
	\caption{\small Time taken and output size for processing logs of 4K \tpcc transactions, while varying admissibility settings of SP-Runs Admitted and Loop-Iters Admitted.
		The most admissible setting (256,8) is \textasciitilde20$\times$ slower than (1,1) and the generated graph is \textasciitilde450$\times$ larger.}
	\label{f:topk-opts}
	\vspace{-1em}
\end{figure}

As discussed in \Cref{s:opts}, several filtering and compression techniques of \sys can speedup the processing of query logs, and reduce the noise of traditional dynamic provenance extraction systems. We evaluate strengths of these techniques next. 

\stitle{Benchmark.} For these experiments, we run \tpcc for a total of 4K \tpcc transactions (16 clients), of which \texttt{new\_order} and \texttt{payment} transactions are invoked 1707 and 1738 times, respectively.
The database emits \textasciitilde250 events/transaction (\textasciitilde1M events overall).

\newcommand{\loopiters}{\textbf{\texttt{Loop-Iters Admitted}}\xspace}
\newcommand{\sprunsadmitted}{\textbf{\texttt{SP-Runs Admitted}}\xspace}

\stitle{Optimizations.} We compare \sys with and without optimizations. Our comparisons are one optimization at a time to better understand the value of the corresponding optimization.  More specifically, recall that the \tpcc statements are executed 100s of times, and some of the executions result in 100s of loop iterations. As such, (1)~loop-compression, (2)~dropping uninteresting activities, and (3)~filters on activities from uninteresting connections are valuable. We denote the loop compression optimization as \loopiters and vary how many iterations are admitted.
The other two filters (i.e., uninteresting activities and filters on activities from uninteresting connections) interoperate in \sys, as we discussed in~\Cref{s:opts}. Thus, we group them, denote the technique as \textbf{\texttt{SP-Runs Admitted}}, and vary how many of the latest runs are admitted (per the last-K semantics of~\Cref{s:opts}). Finally, we aim to see the effect of \textbf{dropping levels of aggregation} and \textbf{events}. (We omit experiments for filters on event metadata and queries for brevity, noting that their overall impact varies based on their selectivity.)

\stitle{Metrics.} We evaluate \sys with and without optimizations based on end-to-end latency and size of provenance graphs emitted. We define the graph size to be the total \#nodes and \#edges.

\stitle{\loopiters and \sprunsadmitted.} The results of our evaluation for these two optimizations are shown in \Cref{f:topk-opts}. The most aggressive setting is (1,1), bottom-left corner tile in \Cref{f:topk-opts},  which prompts \sprunsadmitted to remove all but the latest run of the 5 \tpcc \sproc{}s, and \loopiters to identify unique control flows in an activity and retain the latest iteration. 
For instance, with the setting (1,1), 1706 (out of 1707) \xruntree{}s corresponding to \texttt{new\_order} runs would be dropped, and the latest \xruntree would have 75\% fewer SQL statements.
Consequently, \lineageextractor needs to analyze only 191 SQL statements, thereby reducing end-to-end latency.
Setting the optimization to (256, 8), shown in the top-right corner tile in \Cref{f:topk-heatmap-time}, means an admission of at most 256 runs of \sproc{}s and at most 8 loop iterations. This leads to 110K SQL statements being analyzed by \lineageextractor, which results in a \textasciitilde20$\times$ slowdown over processing 191 statements.

Finally, the top-right tile in \Cref{f:topk-heatmap-size} corresponds to the worst case scenario for the provenance graph sizes (i.e., output of \sys in absence of optimizations). In absolute terms, the output has more than 450K nodes and edges. The large size of the graph is due to modeling of recurring \sproc{}s and SQL statement runs. Such a large graph can be overwhelming for upstream applications. 
In contrast, the aggressive optimization setting (1,1) leads to a concise, noise-free output model, with just over 1K nodes and edges.

\stitle{Drop levels of aggregation.} Using this optimization we configure \sys to report provenance only at aggregate levels (i.e., at stored procedures) and we drop statement runs and their provenance for noise reduction purposes. This leads to an \textasciitilde80\% drop over the provenance graph size of \sys without the optimization. Note that this optimization is applied during aggregation at the Stitcher, because provenance needs to be aggregated anyways, hence does not improve end-to-end latency. 

\stitle{\texttt{Drop Events}.} Recall that this optimization drops events based on database event buffer availability, leading to extracting reduced provenance graphs. \sys with this optimization drops only 3\% of \#nodes and \#edges for the \tpcc workload (over \sys without the optimization). To better understand this result, we also run \qplan with \texttt{Drop Events} on. The drop was \textasciitilde50\%---further highlighting the prohibitive overheads of query plans and \sys's ability to process higher TPM workloads.

\stitle{Provenance querying.} Finally, note that our experiments focus on the problem of provenance capture of~\Cref{s:bg:problem}. We omit analysis on provenance querying since \sys is external to the data catalog that serves provenance queries. We note, however, that in production workloads we have not observed non-interactive query response times (>.5s) over \sys's output. 

\stitle{Takeaways: } Our results demonstrate that our application-aware optimizations lead to substantial noise reduction over provenance graphs (more than two orders of magnitude graph size reduction) and improve extraction performance (\textasciitilde20$\times$ speedup) significantly.

\subsection{Discussion on other systems}
\label{ss:exp:incomparison}

We conclude our discussion with a comparison between \sys and SAC~\cite{tang:2019:sac}, Spline~\cite{scherbaum2018spline}, and OpenLineage for Spark~\cite{url:openlineage}. For these experiments we used the \tpcds benchmark (25 iterations of 99 queries) since these systems target analytical workloads, and we focus our analysis only on the output graph. (As discussed in~\Cref{s:exp-sys-perf}, performance comparisons are not meaningful. This is why we introduced \qplan in~\Cref{s:exp-sys-perf}, to compare \sys with state-of-the-art techniques in a principled way.) 

Overall, \sys supports provenance aggregation, query runtime metadata (CPU and \#records), user and client connection details, inference of static queries, and column-level provenance that SAC, Spline, and OpenLineage for Spark lack.  In contrast, these systems actually embed Spark plans in their output model that \sys does not. Such plans can be large in size and, as such, while \sys outputs more metadata its output is often on par or smaller than the one of the other systems. Furthermore, customer feedback on outputting plans highlighted a disconnect: data governance teams that consume the output of provenance extractors are not DBAs to understand or make use of such plans. Finally, we note that our optimizations further reduce the output size in contrast to SAC, Spline, and OpenLineage that do not provide noise reduction optimizations during extraction (e.g., in our \tpcds workload the output of \sys is \textasciitilde$10\times$ smaller compared to SAC).

\section{Related Work}
\label{s:related}

We describe related work in the areas of metadata and provenance management. To highlight the importance of these domains, we note that several techniques and applications of metadata and provenance management have been included in many high-profile products and open source systems. Adaptive~\cite{url:adaptive}, Alation~\cite{url:alation}, Acryl Data~\cite{url:acryldata}, Alex~\cite{url:alex}, ASG~\cite{url:asg}, Collibra~\cite{url:colibra}, Data Advantage Group~\cite{url:dag}, Datakin~\cite{url:datakin}, data.world~\cite{url:dataworld}, Amazon~\cite{url:datazone}, erwin~\cite{url:erwin}, Global IDs~\cite{url:globalids}, Google~\cite{url:gdc}, IBM~\cite{url:ibm-infosphere}, Informatica~\cite{url:informatica}, Microsoft~\cite{url:purview}, Precisely~\cite{url:precisely}, Semantic Web Company~\cite{url:semanticweb}, Smartlogic~\cite{url:smartlogic}, and Syniti~\cite{url:syniti} are only a few companies with offerings that include metadata and provenance capabilities. Also, open source systems in the same domain areas include Apache Atlas~\cite{url:atlas}, Egeria~\cite{url:egeria-lineage}, OpenLineage~\cite{url:openlineage}, DataHub~\cite{url:datahub}, and Spline~\cite{scherbaum2018spline}, among others.

Metadata management is a sub-field of data management with over five decades of research and practice~\cite{bernstein:2000:modelmanagementvision,nargesian:2019:datalakemanagementtutorial}. Problems of focus in this space include discovery~\cite{DBLP:conf/cidr/DengFAWSEIMO017,DBLP:conf/icde/FernandezAKYMS18,DBLP:journals/debu/MillerNZCPA18}, classification~\cite{10.1145/3209900.3209911,10.14778/3192965.3192973}, extraction~\cite{10.1145/2882903.2899391,10.14778/2994509.2994534}, or storage and querying of metadata~\cite{10.1145/2882903.2903730,DBLP:conf/cidr/HellersteinSGSA17,psallidas2018smoke,widom:2005:trio}. Furthermore, metadata management systems are central for many metadata-driven applications and problems, including: data integration and exchange~\cite{ANADIOTIS2022101846,10.14778/1687627.1687750,DBLP:journals/tcs/FaginKMP05,10.1145/4229.4233,10.1145/1376616.1376702}, schema (and general metadata) evolution~\cite{DBLP:conf/iceis/CurinoMTZ08,10.14778/1453856.1453939}, lifecycle management and versioning~\cite{DBLP:conf/cidr/BhardwajBCDEMP15,10.14778/2824032.2824035,provdb2017miao}, profiling~\cite{DBLP:journals/vldb/AbedjanGN15,DBLP:conf/icde/AbedjanGJN14}, data cleansing~\cite{10.1145/2882903.2899391,10.14778/1952376.1952378,10.5555/1841211}, reproducibility~\cite{10.1145/1142473.1142574,10.14778/3415478.3415556}, enterprise search~\cite{kruschwitz2017searching}, and auditing~\cite{url:ccpa,url:gdpr}. Our work is closely related and largely orthogonal to these lines of work. In particular, the techniques we propose can extract a rich provenance model with a multitude of metadata on queries, query runs, client connections, and datasets (e.g., tables or outputs of ad hoc queries). With such rich information, we can better assist metadata management applications to better drive their logic.

Provenance management is a subfield of data and metadata management with a focus on capturing, modeling, and querying the connections between input and output data elements across a workflow. Traditionally, in the context of databases, provenance is classified into coarse-grained~\cite{amsterdamer2011putting, Cheney2009,herschel2017survey,ludascher2006scientific,hull2006taverna,olston2011ibis,scherbaum2018spline,tang:2019:sac,provdb2017miao} and fine-grained \cite{dbnotes2004bhagwat,Cheney2009,cui2001lineage,glavic:2009:perm,xing:2017:gprom,ikedathesis,green:2007:orchestra,herschel2017survey,psallidas2018smoke,widom:2005:trio,woodruff1997supporting}. The latter encodes the relationships between input and output records or cells, while the former focuses on modeling relationships at a coarse level (e.g., tables and columns). Our proposal is related to capturing dynamic coarse-grained provenance information in the context of database systems. In contrast to prior work, however, we highlighted unique challenges and proposed corresponding techniques to extract semantically rich provenance information from event logs efficiently.

\section{Conclusion}
\label{s:conclusion}

In this paper, we presented \sys, the Microsoft Purview dynamic provenance extraction system over database logs. \sys improves over prior work by processing logs carrying query text, aggregating provenance information, and filtering noise during extraction. We believe our work is a step towards optimized provenance extraction systems, and a pointer towards important future work (e.g., introduce more complicated filtering techniques and extract provenance by processing logs in a distributed fashion to cope with even higher transactional loads).

\balance

\bibliographystyle{ACM-Reference-Format}
\bibliography{oneprovenance-techreport}


\begin{thebibliography}{86}


\ifx \showCODEN    \undefined \def \showCODEN     #1{\unskip}     \fi
\ifx \showDOI      \undefined \def \showDOI       #1{#1}\fi
\ifx \showISBNx    \undefined \def \showISBNx     #1{\unskip}     \fi
\ifx \showISBNxiii \undefined \def \showISBNxiii  #1{\unskip}     \fi
\ifx \showISSN     \undefined \def \showISSN      #1{\unskip}     \fi
\ifx \showLCCN     \undefined \def \showLCCN      #1{\unskip}     \fi
\ifx \shownote     \undefined \def \shownote      #1{#1}          \fi
\ifx \showarticletitle \undefined \def \showarticletitle #1{#1}   \fi
\ifx \showURL      \undefined \def \showURL       {\relax}        \fi
\providecommand\bibfield[2]{#2}
\providecommand\bibinfo[2]{#2}
\providecommand\natexlab[1]{#1}
\providecommand\showeprint[2][]{arXiv:#2}

\bibitem[Abadi et~al\mbox{.}(2022)]%
        {abadi2022seattlereport}
\bibfield{author}{\bibinfo{person}{Daniel Abadi}, \bibinfo{person}{Anastasia
  Ailamaki}, \bibinfo{person}{David Andersen}, \bibinfo{person}{Peter Bailis},
  \bibinfo{person}{Magdalena Balazinska}, \bibinfo{person}{Philip~A.
  Bernstein}, \bibinfo{person}{Peter Boncz}, \bibinfo{person}{Surajit
  Chaudhuri}, \bibinfo{person}{Alvin Cheung}, \bibinfo{person}{Anhai Doan},
  \bibinfo{person}{Luna Dong}, \bibinfo{person}{Michael~J. Franklin},
  \bibinfo{person}{Juliana Freire}, \bibinfo{person}{Alon Halevy},
  \bibinfo{person}{Joseph~M. Hellerstein}, \bibinfo{person}{Stratos Idreos},
  \bibinfo{person}{Donald Kossmann}, \bibinfo{person}{Tim Kraska},
  \bibinfo{person}{Sailesh Krishnamurthy}, \bibinfo{person}{Volker Markl},
  \bibinfo{person}{Sergey Melnik}, \bibinfo{person}{Tova Milo},
  \bibinfo{person}{C. Mohan}, \bibinfo{person}{Thomas Neumann},
  \bibinfo{person}{Beng~Chin Ooi}, \bibinfo{person}{Fatma Ozcan},
  \bibinfo{person}{Jignesh Patel}, \bibinfo{person}{Andrew Pavlo},
  \bibinfo{person}{Raluca Popa}, \bibinfo{person}{Raghu Ramakrishnan},
  \bibinfo{person}{Christopher Re}, \bibinfo{person}{Michael Stonebraker},
  {and} \bibinfo{person}{Dan Suciu}.} \bibinfo{year}{2022}\natexlab{}.
\newblock \showarticletitle{The Seattle Report on Database Research}.
\newblock \bibinfo{journal}{\emph{Commun. ACM}} \bibinfo{volume}{65},
  \bibinfo{number}{8} (\bibinfo{date}{jul} \bibinfo{year}{2022}),
  \bibinfo{pages}{72–79}.
\newblock
\showISSN{0001-0782}
\urldef\tempurl%
\url{https://doi.org/10.1145/3524284}
\showDOI{\tempurl}


\bibitem[Abedjan et~al\mbox{.}(2015)]%
        {DBLP:journals/vldb/AbedjanGN15}
\bibfield{author}{\bibinfo{person}{Ziawasch Abedjan}, \bibinfo{person}{Lukasz
  Golab}, {and} \bibinfo{person}{Felix Naumann}.}
  \bibinfo{year}{2015}\natexlab{}.
\newblock \showarticletitle{Profiling relational data: a survey}.
\newblock \bibinfo{journal}{\emph{{VLDB} J.}} \bibinfo{volume}{24},
  \bibinfo{number}{4} (\bibinfo{year}{2015}), \bibinfo{pages}{557--581}.
\newblock
\urldef\tempurl%
\url{https://doi.org/10.1007/s00778-015-0389-y}
\showDOI{\tempurl}


\bibitem[Abedjan et~al\mbox{.}(2014)]%
        {DBLP:conf/icde/AbedjanGJN14}
\bibfield{author}{\bibinfo{person}{Ziawasch Abedjan}, \bibinfo{person}{Toni
  Gr{\"{u}}tze}, \bibinfo{person}{Anja Jentzsch}, {and} \bibinfo{person}{Felix
  Naumann}.} \bibinfo{year}{2014}\natexlab{}.
\newblock \showarticletitle{Profiling and mining {RDF} data with ProLOD++}. In
  \bibinfo{booktitle}{\emph{{IEEE} 30th International Conference on Data
  Engineering, Chicago, {ICDE} 2014, IL, USA, March 31 - April 4, 2014}},
  \bibfield{editor}{\bibinfo{person}{Isabel~F. Cruz}, \bibinfo{person}{Elena
  Ferrari}, \bibinfo{person}{Yufei Tao}, \bibinfo{person}{Elisa Bertino}, {and}
  \bibinfo{person}{Goce Trajcevski}} (Eds.). \bibinfo{publisher}{{IEEE}
  Computer Society}, \bibinfo{pages}{1198--1201}.
\newblock
\urldef\tempurl%
\url{https://doi.org/10.1109/ICDE.2014.6816740}
\showDOI{\tempurl}


\bibitem[acryldata(2023)]%
        {url:acryldata}
acryldata \bibinfo{year}{2023}\natexlab{}.
\newblock \bibinfo{title}{{Acryl Data}}.
\newblock \bibinfo{howpublished}{{\small\url{https://www.acryldata.io/}}}.
\newblock


\bibitem[adaptive(2022)]%
        {url:adaptive}
adaptive \bibinfo{year}{2022}\natexlab{}.
\newblock \bibinfo{title}{{Adaptive}}.
\newblock \bibinfo{howpublished}{{\small\url{https://adaptive.com}}}.
\newblock


\bibitem[alation(2022)]%
        {url:alation}
alation \bibinfo{year}{2022}\natexlab{}.
\newblock \bibinfo{title}{{Alation}}.
\newblock \bibinfo{howpublished}{{\small\url{https://alation.com}}}.
\newblock


\bibitem[alex(2022)]%
        {url:alex}
alex \bibinfo{year}{2022}\natexlab{}.
\newblock \bibinfo{title}{{Alex Solutions}}.
\newblock \bibinfo{howpublished}{{\small\url{https://alexsolutions.com.au}}}.
\newblock


\bibitem[Amsterdamer et~al\mbox{.}(2011)]%
        {amsterdamer2011putting}
\bibfield{author}{\bibinfo{person}{Yael Amsterdamer}, \bibinfo{person}{Susan~B
  Davidson}, \bibinfo{person}{Daniel Deutch}, \bibinfo{person}{Tova Milo},
  \bibinfo{person}{Julia Stoyanovich}, {and} \bibinfo{person}{Val Tannen}.}
  \bibinfo{year}{2011}\natexlab{}.
\newblock \showarticletitle{Putting lipstick on pig: Enabling database-style
  workflow provenance}.
\newblock \bibinfo{journal}{\emph{arXiv preprint arXiv:1201.0231}}
  (\bibinfo{year}{2011}).
\newblock


\bibitem[Anadiotis et~al\mbox{.}(2022)]%
        {ANADIOTIS2022101846}
\bibfield{author}{\bibinfo{person}{Angelos~Christos Anadiotis},
  \bibinfo{person}{Oana Balalau}, \bibinfo{person}{Catarina Conceição},
  \bibinfo{person}{Helena Galhardas}, \bibinfo{person}{Mhd~Yamen Haddad},
  \bibinfo{person}{Ioana Manolescu}, \bibinfo{person}{Tayeb Merabti}, {and}
  \bibinfo{person}{Jingmao You}.} \bibinfo{year}{2022}\natexlab{}.
\newblock \showarticletitle{Graph integration of structured, semistructured and
  unstructured data for data journalism}.
\newblock \bibinfo{journal}{\emph{Information Systems}}  \bibinfo{volume}{104}
  (\bibinfo{year}{2022}), \bibinfo{pages}{101846}.
\newblock
\showISSN{0306-4379}
\urldef\tempurl%
\url{https://doi.org/10.1016/j.is.2021.101846}
\showDOI{\tempurl}


\bibitem[asg(2022)]%
        {url:asg}
asg \bibinfo{year}{2022}\natexlab{}.
\newblock \bibinfo{title}{{ASG}}.
\newblock \bibinfo{howpublished}{{\small\url{https://www.asg.com}}}.
\newblock


\bibitem[atlas(2019)]%
        {url:atlas}
atlas \bibinfo{year}{2019}\natexlab{}.
\newblock \bibinfo{title}{{Apache Atlas - Type System}}.
\newblock \bibinfo{howpublished}{{\small
  \url{https://atlas.apache.org/\#/TypeSystem}}}.
\newblock


\bibitem[awsdatazone(2022)]%
        {url:datazone}
awsdatazone \bibinfo{year}{2022}\natexlab{}.
\newblock \bibinfo{title}{{AWS DataZone}}.
\newblock
  \bibinfo{howpublished}{{\small\url{https://aws.amazon.com/datazone/}}}.
\newblock


\bibitem[azurestorage(2022)]%
        {url:azurestorage}
azurestorage \bibinfo{year}{2022}\natexlab{}.
\newblock \bibinfo{title}{{Azure Storage}}.
\newblock
  \bibinfo{howpublished}{{\small\url{https://azure.microsoft.com/services/storage}}}.
\newblock


\bibitem[Bernstein et~al\mbox{.}(1979)]%
        {bernstein:1979:serializability}
\bibfield{author}{\bibinfo{person}{P.A. Bernstein}, \bibinfo{person}{D.W.
  Shipman}, {and} \bibinfo{person}{W.S. Wong}.}
  \bibinfo{year}{1979}\natexlab{}.
\newblock \showarticletitle{Formal Aspects of Serializability in Database
  Concurrency Control}.
\newblock \bibinfo{journal}{\emph{IEEE Transactions on Software Engineering}}
  \bibinfo{volume}{SE-5}, \bibinfo{number}{3} (\bibinfo{year}{1979}),
  \bibinfo{pages}{203--216}.
\newblock


\bibitem[Bernstein et~al\mbox{.}(2000)]%
        {bernstein:2000:modelmanagementvision}
\bibfield{author}{\bibinfo{person}{Phillip~A. Bernstein},
  \bibinfo{person}{Alon~Y. Halevy}, {and} \bibinfo{person}{Rachel~A.
  Pottinger}.} \bibinfo{year}{2000}\natexlab{}.
\newblock \showarticletitle{A Vision for Management of Complex Models}.
\newblock \bibinfo{journal}{\emph{SIGMOD Rec.}} \bibinfo{volume}{29},
  \bibinfo{number}{4} (\bibinfo{date}{Dec.} \bibinfo{year}{2000}),
  \bibinfo{pages}{55–63}.
\newblock
\showISSN{0163-5808}
\urldef\tempurl%
\url{https://doi.org/10.1145/369275.369289}
\showDOI{\tempurl}


\bibitem[Bhagwat et~al\mbox{.}(2004)]%
        {dbnotes2004bhagwat}
\bibfield{author}{\bibinfo{person}{Deepavali Bhagwat}, \bibinfo{person}{Laura
  Chiticariu}, \bibinfo{person}{Wang~Chiew Tan}, {and} \bibinfo{person}{Gaurav
  Vijayvargiya}.} \bibinfo{year}{2004}\natexlab{}.
\newblock \showarticletitle{An Annotation Management System for Relational
  Databases}. In \bibinfo{booktitle}{\emph{VLDB}}. \bibinfo{pages}{900--911}.
\newblock


\bibitem[Bhardwaj et~al\mbox{.}(2015)]%
        {DBLP:conf/cidr/BhardwajBCDEMP15}
\bibfield{author}{\bibinfo{person}{Anant~P. Bhardwaj}, \bibinfo{person}{Souvik
  Bhattacherjee}, \bibinfo{person}{Amit Chavan}, \bibinfo{person}{Amol
  Deshpande}, \bibinfo{person}{Aaron~J. Elmore}, \bibinfo{person}{Samuel
  Madden}, {and} \bibinfo{person}{Aditya~G. Parameswaran}.}
  \bibinfo{year}{2015}\natexlab{}.
\newblock \showarticletitle{DataHub: Collaborative Data Science {\&} Dataset
  Version Management at Scale}. In \bibinfo{booktitle}{\emph{Seventh Biennial
  Conference on Innovative Data Systems Research, {CIDR} 2015, Asilomar, CA,
  USA, January 4-7, 2015, Online Proceedings}}.
  \bibinfo{publisher}{www.cidrdb.org}.
\newblock
\urldef\tempurl%
\url{http://cidrdb.org/cidr2015/Papers/CIDR15\_Paper18.pdf}
\showURL{%
\tempurl}


\bibitem[Bhattacherjee et~al\mbox{.}(2015)]%
        {10.14778/2824032.2824035}
\bibfield{author}{\bibinfo{person}{Souvik Bhattacherjee}, \bibinfo{person}{Amit
  Chavan}, \bibinfo{person}{Silu Huang}, \bibinfo{person}{Amol Deshpande},
  {and} \bibinfo{person}{Aditya Parameswaran}.}
  \bibinfo{year}{2015}\natexlab{}.
\newblock \showarticletitle{Principles of Dataset Versioning: Exploring the
  Recreation/Storage Tradeoff}.
\newblock \bibinfo{journal}{\emph{Proc. VLDB Endow.}} \bibinfo{volume}{8},
  \bibinfo{number}{12} (\bibinfo{date}{aug} \bibinfo{year}{2015}),
  \bibinfo{pages}{1346–1357}.
\newblock
\showISSN{2150-8097}
\urldef\tempurl%
\url{https://doi.org/10.14778/2824032.2824035}
\showDOI{\tempurl}


\bibitem[Brackenbury et~al\mbox{.}(2018)]%
        {10.1145/3209900.3209911}
\bibfield{author}{\bibinfo{person}{Will Brackenbury}, \bibinfo{person}{Rui
  Liu}, \bibinfo{person}{Mainack Mondal}, \bibinfo{person}{Aaron~J. Elmore},
  \bibinfo{person}{Blase Ur}, \bibinfo{person}{Kyle Chard}, {and}
  \bibinfo{person}{Michael~J. Franklin}.} \bibinfo{year}{2018}\natexlab{}.
\newblock \showarticletitle{Draining the Data Swamp: A Similarity-Based
  Approach}. In \bibinfo{booktitle}{\emph{Proceedings of the Workshop on
  Human-In-the-Loop Data Analytics}} (Houston, TX, USA)
  \emph{(\bibinfo{series}{HILDA'18})}. \bibinfo{publisher}{Association for
  Computing Machinery}, \bibinfo{address}{New York, NY, USA}, Article
  \bibinfo{articleno}{13}, \bibinfo{numpages}{7}~pages.
\newblock
\showISBNx{9781450358279}
\urldef\tempurl%
\url{https://doi.org/10.1145/3209900.3209911}
\showDOI{\tempurl}


\bibitem[Cafarella et~al\mbox{.}(2009)]%
        {10.14778/1687627.1687750}
\bibfield{author}{\bibinfo{person}{Michael~J. Cafarella}, \bibinfo{person}{Alon
  Halevy}, {and} \bibinfo{person}{Nodira Khoussainova}.}
  \bibinfo{year}{2009}\natexlab{}.
\newblock \showarticletitle{Data Integration for the Relational Web}.
\newblock \bibinfo{journal}{\emph{Proc. VLDB Endow.}} \bibinfo{volume}{2},
  \bibinfo{number}{1} (\bibinfo{date}{aug} \bibinfo{year}{2009}),
  \bibinfo{pages}{1090–1101}.
\newblock
\showISSN{2150-8097}
\urldef\tempurl%
\url{https://doi.org/10.14778/1687627.1687750}
\showDOI{\tempurl}


\bibitem[Callahan et~al\mbox{.}(2006)]%
        {10.1145/1142473.1142574}
\bibfield{author}{\bibinfo{person}{Steven~P. Callahan},
  \bibinfo{person}{Juliana Freire}, \bibinfo{person}{Emanuele Santos},
  \bibinfo{person}{Carlos~E. Scheidegger}, \bibinfo{person}{Cl\'{a}udio~T.
  Silva}, {and} \bibinfo{person}{Huy~T. Vo}.} \bibinfo{year}{2006}\natexlab{}.
\newblock \showarticletitle{VisTrails: Visualization Meets Data Management}. In
  \bibinfo{booktitle}{\emph{Proceedings of the 2006 ACM SIGMOD International
  Conference on Management of Data}} (Chicago, IL, USA)
  \emph{(\bibinfo{series}{SIGMOD '06})}. \bibinfo{publisher}{Association for
  Computing Machinery}, \bibinfo{address}{New York, NY, USA},
  \bibinfo{pages}{745–747}.
\newblock
\showISBNx{1595934340}
\urldef\tempurl%
\url{https://doi.org/10.1145/1142473.1142574}
\showDOI{\tempurl}


\bibitem[ccpa(2022)]%
        {url:ccpa}
ccpa \bibinfo{year}{2022}\natexlab{}.
\newblock \bibinfo{title}{{California Consumer Privacy Act (CCPA)}}.
\newblock
  \bibinfo{howpublished}{{\small\url{https://oag.ca.gov/privacy/ccpa}}}.
\newblock


\bibitem[Cheney et~al\mbox{.}(2009)]%
        {Cheney2009}
\bibfield{author}{\bibinfo{person}{James Cheney}, \bibinfo{person}{Laura
  Chiticariu}, {and} \bibinfo{person}{Wang~Chiew Tan}.}
  \bibinfo{year}{2009}\natexlab{}.
\newblock \showarticletitle{Provenance in databases: Why, how, and where}.
\newblock \bibinfo{journal}{\emph{{Foundations and Trends{\textregistered}~in
  Databases}}} \bibinfo{volume}{1}, \bibinfo{number}{4} (\bibinfo{year}{2009}),
  \bibinfo{pages}{379--474}.
\newblock


\bibitem[colibra(2022)]%
        {url:colibra}
colibra \bibinfo{year}{2022}\natexlab{}.
\newblock \bibinfo{title}{{Colibra}}.
\newblock \bibinfo{howpublished}{{\small\url{https://colibra.com}}}.
\newblock


\bibitem[createeventsession(2022)]%
        {url:createeventsession}
createeventsession \bibinfo{year}{2022}\natexlab{}.
\newblock \bibinfo{title}{{Create Event Session}}.
\newblock
  \bibinfo{howpublished}{{\small\url{https://docs.microsoft.com/en-us/sql/t-sql/statements/create-event-session-transact-sql?view=sql-server-ver15}}}.
\newblock


\bibitem[Cui(2001)]%
        {cui2001lineage}
\bibfield{author}{\bibinfo{person}{Yingwei Cui}.}
  \bibinfo{year}{2001}\natexlab{}.
\newblock \emph{\bibinfo{title}{Lineage tracing in data warehouses}}.
\newblock \bibinfo{thesistype}{Ph.\,D. Dissertation}. \bibinfo{school}{Stanford
  University}.
\newblock


\bibitem[Curino et~al\mbox{.}(2008b)]%
        {DBLP:conf/iceis/CurinoMTZ08}
\bibfield{author}{\bibinfo{person}{Carlo Curino}, \bibinfo{person}{Hyun~Jin
  Moon}, \bibinfo{person}{Letizia Tanca}, {and} \bibinfo{person}{Carlo
  Zaniolo}.} \bibinfo{year}{2008}\natexlab{b}.
\newblock \showarticletitle{Schema Evolution in Wikipedia - Toward a Web
  Information System Benchmark}. In \bibinfo{booktitle}{\emph{{ICEIS} 2008 -
  Proceedings of the Tenth International Conference on Enterprise Information
  Systems, Volume DISI, Barcelona, Spain, June 12-16, 2008}},
  \bibfield{editor}{\bibinfo{person}{Jos{\'{e}} Cordeiro} {and}
  \bibinfo{person}{Joaquim Filipe}} (Eds.). \bibinfo{pages}{323--332}.
\newblock


\bibitem[Curino et~al\mbox{.}(2008a)]%
        {10.14778/1453856.1453939}
\bibfield{author}{\bibinfo{person}{Carlo~A. Curino}, \bibinfo{person}{Hyun~J.
  Moon}, {and} \bibinfo{person}{Carlo Zaniolo}.}
  \bibinfo{year}{2008}\natexlab{a}.
\newblock \showarticletitle{Graceful Database Schema Evolution: The PRISM
  Workbench}.
\newblock \bibinfo{journal}{\emph{Proc. VLDB Endow.}} \bibinfo{volume}{1},
  \bibinfo{number}{1} (\bibinfo{date}{aug} \bibinfo{year}{2008}),
  \bibinfo{pages}{761–772}.
\newblock
\showISSN{2150-8097}
\urldef\tempurl%
\url{https://doi.org/10.14778/1453856.1453939}
\showDOI{\tempurl}


\bibitem[dag(2022)]%
        {url:dag}
dag \bibinfo{year}{2022}\natexlab{}.
\newblock \bibinfo{title}{{Data Advantage Group}}.
\newblock \bibinfo{howpublished}{{\small\url{https://www.dag.com}}}.
\newblock


\bibitem[Das~Sarma et~al\mbox{.}(2008)]%
        {10.1145/1376616.1376702}
\bibfield{author}{\bibinfo{person}{Anish Das~Sarma}, \bibinfo{person}{Xin
  Dong}, {and} \bibinfo{person}{Alon Halevy}.} \bibinfo{year}{2008}\natexlab{}.
\newblock \showarticletitle{Bootstrapping Pay-as-You-Go Data Integration
  Systems}. In \bibinfo{booktitle}{\emph{Proceedings of the 2008 ACM SIGMOD
  International Conference on Management of Data}} (Vancouver, Canada)
  \emph{(\bibinfo{series}{SIGMOD '08})}. \bibinfo{publisher}{Association for
  Computing Machinery}, \bibinfo{address}{New York, NY, USA},
  \bibinfo{pages}{861–874}.
\newblock
\showISBNx{9781605581026}
\urldef\tempurl%
\url{https://doi.org/10.1145/1376616.1376702}
\showDOI{\tempurl}


\bibitem[datahub(2023)]%
        {url:datahub}
datahub \bibinfo{year}{2023}\natexlab{}.
\newblock \bibinfo{title}{{DataHub}}.
\newblock \bibinfo{howpublished}{{\small\url{https://datahubproject.io/}}}.
\newblock


\bibitem[datakin(2022)]%
        {url:datakin}
datakin \bibinfo{year}{2022}\natexlab{}.
\newblock \bibinfo{title}{{Datakin}}.
\newblock \bibinfo{howpublished}{{\small\url{https://datakin.com}}}.
\newblock


\bibitem[dataworld(2022)]%
        {url:dataworld}
dataworld \bibinfo{year}{2022}\natexlab{}.
\newblock \bibinfo{title}{{data.world}}.
\newblock \bibinfo{howpublished}{{\small\url{https://data.world}}}.
\newblock


\bibitem[Deng et~al\mbox{.}(2017)]%
        {DBLP:conf/cidr/DengFAWSEIMO017}
\bibfield{author}{\bibinfo{person}{Dong Deng}, \bibinfo{person}{Raul~Castro
  Fernandez}, \bibinfo{person}{Ziawasch Abedjan}, \bibinfo{person}{Sibo Wang},
  \bibinfo{person}{Michael Stonebraker}, \bibinfo{person}{Ahmed~K. Elmagarmid},
  \bibinfo{person}{Ihab~F. Ilyas}, \bibinfo{person}{Samuel Madden},
  \bibinfo{person}{Mourad Ouzzani}, {and} \bibinfo{person}{Nan Tang}.}
  \bibinfo{year}{2017}\natexlab{}.
\newblock \showarticletitle{The Data Civilizer System}. In
  \bibinfo{booktitle}{\emph{8th Biennial Conference on Innovative Data Systems
  Research, {CIDR} 2017, Chaminade, CA, USA, January 8-11, 2017, Online
  Proceedings}}. \bibinfo{publisher}{www.cidrdb.org}.
\newblock
\urldef\tempurl%
\url{http://cidrdb.org/cidr2017/papers/p44-deng-cidr17.pdf}
\showURL{%
\tempurl}


\bibitem[egeria-lineage(2022a)]%
        {url:egeria-lineage}
egeria-lineage \bibinfo{year}{2022}\natexlab{a}.
\newblock \bibinfo{title}{{Egeria - Lineage Management}}.
\newblock \bibinfo{howpublished}{{\small
  \url{https://egeria-project.org/features/lineage-management/overview/\#lineage-styles}}}.
\newblock


\bibitem[egeria-lineage(2022b)]%
        {url:openlineage}
egeria-lineage \bibinfo{year}{2022}\natexlab{b}.
\newblock \bibinfo{title}{{OpenLineage - Lineage Management}}.
\newblock \bibinfo{howpublished}{{\small \url{https://openlineage.io/}}}.
\newblock


\bibitem[erwin(2022)]%
        {url:erwin}
erwin \bibinfo{year}{2022}\natexlab{}.
\newblock \bibinfo{title}{{erwin}}.
\newblock \bibinfo{howpublished}{{\small\url{https://www.erwin.com}}}.
\newblock


\bibitem[Fagin et~al\mbox{.}(2005)]%
        {DBLP:journals/tcs/FaginKMP05}
\bibfield{author}{\bibinfo{person}{Ronald Fagin}, \bibinfo{person}{Phokion~G.
  Kolaitis}, \bibinfo{person}{Ren{\'{e}}e~J. Miller}, {and}
  \bibinfo{person}{Lucian Popa}.} \bibinfo{year}{2005}\natexlab{}.
\newblock \showarticletitle{Data exchange: semantics and query answering}.
\newblock \bibinfo{journal}{\emph{Theor. Comput. Sci.}} \bibinfo{volume}{336},
  \bibinfo{number}{1} (\bibinfo{year}{2005}), \bibinfo{pages}{89--124}.
\newblock
\urldef\tempurl%
\url{https://doi.org/10.1016/j.tcs.2004.10.033}
\showDOI{\tempurl}


\bibitem[Farid et~al\mbox{.}(2016)]%
        {10.1145/2882903.2899391}
\bibfield{author}{\bibinfo{person}{Mina Farid}, \bibinfo{person}{Alexandra
  Roatis}, \bibinfo{person}{Ihab~F. Ilyas}, \bibinfo{person}{Hella-Franziska
  Hoffmann}, {and} \bibinfo{person}{Xu Chu}.} \bibinfo{year}{2016}\natexlab{}.
\newblock \showarticletitle{CLAMS: Bringing Quality to Data Lakes}. In
  \bibinfo{booktitle}{\emph{Proceedings of the 2016 International Conference on
  Management of Data}} (San Francisco, California, USA)
  \emph{(\bibinfo{series}{SIGMOD '16})}. \bibinfo{publisher}{Association for
  Computing Machinery}, \bibinfo{address}{New York, NY, USA},
  \bibinfo{pages}{2089–2092}.
\newblock
\showISBNx{9781450335317}
\urldef\tempurl%
\url{https://doi.org/10.1145/2882903.2899391}
\showDOI{\tempurl}


\bibitem[Fernandez et~al\mbox{.}(2018)]%
        {DBLP:conf/icde/FernandezAKYMS18}
\bibfield{author}{\bibinfo{person}{Raul~Castro Fernandez},
  \bibinfo{person}{Ziawasch Abedjan}, \bibinfo{person}{Famien Koko},
  \bibinfo{person}{Gina Yuan}, \bibinfo{person}{Samuel Madden}, {and}
  \bibinfo{person}{Michael Stonebraker}.} \bibinfo{year}{2018}\natexlab{}.
\newblock \showarticletitle{Aurum: {A} Data Discovery System}. In
  \bibinfo{booktitle}{\emph{34th {IEEE} International Conference on Data
  Engineering, {ICDE} 2018, Paris, France, April 16-19, 2018}}.
  \bibinfo{publisher}{{IEEE} Computer Society}, \bibinfo{pages}{1001--1012}.
\newblock
\urldef\tempurl%
\url{https://doi.org/10.1109/ICDE.2018.00094}
\showDOI{\tempurl}


\bibitem[gartner-governance(2020)]%
        {url:gartner:governance}
gartner-governance \bibinfo{year}{2020}\natexlab{}.
\newblock \bibinfo{title}{{Gartner Report on Metadata Management Solutions}}.
\newblock \bibinfo{howpublished}{{\small
  \url{https://www.gartner.com/en/documents/3993025}}}.
\newblock


\bibitem[gdc(2022)]%
        {url:gdc}
gdc \bibinfo{year}{2022}\natexlab{}.
\newblock \bibinfo{title}{{Google Data Catalog}}.
\newblock
  \bibinfo{howpublished}{{\small\url{https://cloud.google.com/data-catalog}}}.
\newblock


\bibitem[gdpr(2022)]%
        {url:gdpr}
gdpr \bibinfo{year}{2022}\natexlab{}.
\newblock \bibinfo{title}{{General Data Protection Regulation (EU GDPR)}}.
\newblock \bibinfo{howpublished}{{\small\url{https://gdpr-info.eu}}}.
\newblock


\bibitem[Glavic and Alonso(2009)]%
        {glavic:2009:perm}
\bibfield{author}{\bibinfo{person}{Boris Glavic} {and} \bibinfo{person}{Gustavo
  Alonso}.} \bibinfo{year}{2009}\natexlab{}.
\newblock \showarticletitle{Perm: Processing provenance and data on the same
  data model through query rewriting}. In \bibinfo{booktitle}{\emph{ICDE}}.
\newblock


\bibitem[globalids(2022)]%
        {url:globalids}
globalids \bibinfo{year}{2022}\natexlab{}.
\newblock \bibinfo{title}{{Global IDs}}.
\newblock \bibinfo{howpublished}{{\small\url{https://www.globalids.com}}}.
\newblock


\bibitem[Granzen(2021)]%
        {url:forrester:governance}
\bibfield{author}{\bibinfo{person}{Achim Granzen}.}
  \bibinfo{year}{2021}\natexlab{}.
\newblock \bibinfo{title}{Data Governance Solutions}.
\newblock \bibinfo{howpublished}{{\small
  \url{https://www.forrester.com/report/the-forrester-wave-tm-data-governance-solutions-q3-2021/RES161533}}}.
\newblock


\bibitem[Green et~al\mbox{.}(2007)]%
        {green:2007:orchestra}
\bibfield{author}{\bibinfo{person}{Todd~J. Green}, \bibinfo{person}{Grigoris
  Karvounarakis}, \bibinfo{person}{Zachary~G. Ives}, {and} \bibinfo{person}{Val
  Tannen}.} \bibinfo{year}{2007}\natexlab{}.
\newblock \showarticletitle{Update Exchange with Mappings and Provenance}. In
  \bibinfo{booktitle}{\emph{VLDB}}.
\newblock


\bibitem[Gupta and Ramachandra(2021)]%
        {procbench2021}
\bibfield{author}{\bibinfo{person}{Surabhi Gupta} {and}
  \bibinfo{person}{Karthik Ramachandra}.} \bibinfo{year}{2021}\natexlab{}.
\newblock \showarticletitle{Procedural Extensions of SQL: Understanding Their
  Usage in the Wild}.
\newblock \bibinfo{journal}{\emph{Proc. VLDB Endow.}} \bibinfo{volume}{14},
  \bibinfo{number}{8} (\bibinfo{date}{apr} \bibinfo{year}{2021}),
  \bibinfo{pages}{1378–1391}.
\newblock
\showISSN{2150-8097}
\urldef\tempurl%
\url{https://doi.org/10.14778/3457390.3457402}
\showDOI{\tempurl}


\bibitem[Halevy et~al\mbox{.}(2016)]%
        {10.1145/2882903.2903730}
\bibfield{author}{\bibinfo{person}{Alon Halevy}, \bibinfo{person}{Flip Korn},
  \bibinfo{person}{Natalya~F. Noy}, \bibinfo{person}{Christopher Olston},
  \bibinfo{person}{Neoklis Polyzotis}, \bibinfo{person}{Sudip Roy}, {and}
  \bibinfo{person}{Steven~Euijong Whang}.} \bibinfo{year}{2016}\natexlab{}.
\newblock \showarticletitle{Goods: Organizing Google's Datasets}. In
  \bibinfo{booktitle}{\emph{Proceedings of the 2016 International Conference on
  Management of Data}} (San Francisco, California, USA)
  \emph{(\bibinfo{series}{SIGMOD '16})}. \bibinfo{publisher}{Association for
  Computing Machinery}, \bibinfo{address}{New York, NY, USA},
  \bibinfo{pages}{795–806}.
\newblock
\showISBNx{9781450335317}
\urldef\tempurl%
\url{https://doi.org/10.1145/2882903.2903730}
\showDOI{\tempurl}


\bibitem[hammerdb(2022)]%
        {url:hammerdb:website}
hammerdb \bibinfo{year}{2022}\natexlab{}.
\newblock \bibinfo{title}{{HammerDB}}.
\newblock \bibinfo{howpublished}{{\small\url{https://www.hammerdb.com}}}.
\newblock


\bibitem[hammerdbgithub(2022)]%
        {url:hammerdb:github}
hammerdbgithub \bibinfo{year}{2022}\natexlab{}.
\newblock \bibinfo{title}{{HammerDB on GitHub}}.
\newblock
  \bibinfo{howpublished}{{\small\url{https://github.com/TPC-Council/HammerDB}}}.
\newblock


\bibitem[Heimbigner and McLeod(1985)]%
        {10.1145/4229.4233}
\bibfield{author}{\bibinfo{person}{Dennis Heimbigner} {and}
  \bibinfo{person}{Dennis McLeod}.} \bibinfo{year}{1985}\natexlab{}.
\newblock \showarticletitle{A Federated Architecture for Information
  Management}.
\newblock \bibinfo{journal}{\emph{ACM Trans. Inf. Syst.}} \bibinfo{volume}{3},
  \bibinfo{number}{3} (\bibinfo{date}{jul} \bibinfo{year}{1985}),
  \bibinfo{pages}{253–278}.
\newblock
\showISSN{1046-8188}
\urldef\tempurl%
\url{https://doi.org/10.1145/4229.4233}
\showDOI{\tempurl}


\bibitem[Hellerstein et~al\mbox{.}(2017)]%
        {DBLP:conf/cidr/HellersteinSGSA17}
\bibfield{author}{\bibinfo{person}{Joseph~M. Hellerstein},
  \bibinfo{person}{Vikram Sreekanti}, \bibinfo{person}{Joseph~E. Gonzalez},
  \bibinfo{person}{James Dalton}, \bibinfo{person}{Akon Dey},
  \bibinfo{person}{Sreyashi Nag}, \bibinfo{person}{Krishna Ramachandran},
  \bibinfo{person}{Sudhanshu Arora}, \bibinfo{person}{Arka Bhattacharyya},
  \bibinfo{person}{Shirshanka Das}, \bibinfo{person}{Mark Donsky},
  \bibinfo{person}{Gabriel Fierro}, \bibinfo{person}{Chang She},
  \bibinfo{person}{Carl Steinbach}, \bibinfo{person}{Venkat Subramanian}, {and}
  \bibinfo{person}{Eric Sun}.} \bibinfo{year}{2017}\natexlab{}.
\newblock \showarticletitle{Ground: {A} Data Context Service}. In
  \bibinfo{booktitle}{\emph{8th Biennial Conference on Innovative Data Systems
  Research, {CIDR} 2017, Chaminade, CA, USA, January 8-11, 2017, Online
  Proceedings}}. \bibinfo{publisher}{www.cidrdb.org}.
\newblock
\urldef\tempurl%
\url{http://cidrdb.org/cidr2017/papers/p111-hellerstein-cidr17.pdf}
\showURL{%
\tempurl}


\bibitem[Herschel et~al\mbox{.}(2017)]%
        {herschel2017survey}
\bibfield{author}{\bibinfo{person}{Melanie Herschel}, \bibinfo{person}{Ralf
  Diestelk{\"a}mper}, {and} \bibinfo{person}{Houssem~Ben Lahmar}.}
  \bibinfo{year}{2017}\natexlab{}.
\newblock \showarticletitle{A survey on provenance: What for? What form? What
  from?}
\newblock \bibinfo{journal}{\emph{The VLDB Journal}} \bibinfo{volume}{26},
  \bibinfo{number}{6} (\bibinfo{year}{2017}), \bibinfo{pages}{881--906}.
\newblock


\bibitem[Hull et~al\mbox{.}(2006)]%
        {hull2006taverna}
\bibfield{author}{\bibinfo{person}{Duncan Hull}, \bibinfo{person}{Katy
  Wolstencroft}, \bibinfo{person}{Robert Stevens}, \bibinfo{person}{Carole
  Goble}, \bibinfo{person}{Mathew~R Pocock}, \bibinfo{person}{Peter Li}, {and}
  \bibinfo{person}{Tom Oinn}.} \bibinfo{year}{2006}\natexlab{}.
\newblock \showarticletitle{Taverna: a tool for building and running workflows
  of services}.
\newblock \bibinfo{journal}{\emph{Nucleic acids research}}
  \bibinfo{volume}{34}, \bibinfo{number}{suppl\_2} (\bibinfo{year}{2006}),
  \bibinfo{pages}{W729--W732}.
\newblock


\bibitem[Ikeda(2012)]%
        {ikedathesis}
\bibfield{author}{\bibinfo{person}{Robert Ikeda}.}
  \bibinfo{year}{2012}\natexlab{}.
\newblock \emph{\bibinfo{title}{Provenance In Data-Oriented Workflows}}.
\newblock \bibinfo{thesistype}{Ph.\,D. Dissertation}. \bibinfo{school}{Stanford
  University}.
\newblock


\bibitem[informatica(2022)]%
        {url:informatica}
informatica \bibinfo{year}{2022}\natexlab{}.
\newblock \bibinfo{title}{{Informatica}}.
\newblock \bibinfo{howpublished}{{\small\url{https://www.informatica.com}}}.
\newblock


\bibitem[infosphere(2022)]%
        {url:ibm-infosphere}
infosphere \bibinfo{year}{2022}\natexlab{}.
\newblock \bibinfo{title}{{IBM Infosphere}}.
\newblock
  \bibinfo{howpublished}{{\small\url{https://www.ibm.com/analytics/information-server}}}.
\newblock


\bibitem[Kruschwitz et~al\mbox{.}(2017)]%
        {kruschwitz2017searching}
\bibfield{author}{\bibinfo{person}{Udo Kruschwitz}, \bibinfo{person}{Charlie
  Hull}, {et~al\mbox{.}}} \bibinfo{year}{2017}\natexlab{}.
\newblock \bibinfo{booktitle}{\emph{Searching the enterprise}}.
  Vol.~\bibinfo{volume}{11}.
\newblock \bibinfo{publisher}{Now Publishers}.
\newblock


\bibitem[Lud{\"a}scher et~al\mbox{.}(2006)]%
        {ludascher2006scientific}
\bibfield{author}{\bibinfo{person}{Bertram Lud{\"a}scher},
  \bibinfo{person}{Ilkay Altintas}, \bibinfo{person}{Chad Berkley},
  \bibinfo{person}{Dan Higgins}, \bibinfo{person}{Efrat Jaeger},
  \bibinfo{person}{Matthew Jones}, \bibinfo{person}{Edward~A Lee},
  \bibinfo{person}{Jing Tao}, {and} \bibinfo{person}{Yang Zhao}.}
  \bibinfo{year}{2006}\natexlab{}.
\newblock \showarticletitle{Scientific workflow management and the Kepler
  system}.
\newblock \bibinfo{journal}{\emph{Concurrency and computation: Practice and
  experience}} \bibinfo{volume}{18}, \bibinfo{number}{10}
  (\bibinfo{year}{2006}), \bibinfo{pages}{1039--1065}.
\newblock


\bibitem[Mavlyutov et~al\mbox{.}(2017)]%
        {mavlyutov2017dependency}
\bibfield{author}{\bibinfo{person}{Ruslan Mavlyutov}, \bibinfo{person}{Carlo
  Curino}, \bibinfo{person}{Boris Asipov}, {and} \bibinfo{person}{Philippe
  Cudr{\'e}-Mauroux}.} \bibinfo{year}{2017}\natexlab{}.
\newblock \showarticletitle{Dependency-Driven Analytics: A Compass for
  Uncharted Data Oceans.}. In \bibinfo{booktitle}{\emph{CIDR}}.
\newblock


\bibitem[Miao et~al\mbox{.}(2017)]%
        {provdb2017miao}
\bibfield{author}{\bibinfo{person}{Hui Miao}, \bibinfo{person}{Amit Chavan},
  {and} \bibinfo{person}{Amol Deshpande}.} \bibinfo{year}{2017}\natexlab{}.
\newblock \showarticletitle{ProvDB: Lifecycle Management of Collaborative
  Analysis Workflows}. In \bibinfo{booktitle}{\emph{Proceedings of the 2nd
  Workshop on Human-In-the-Loop Data Analytics}} (Chicago, IL, USA)
  \emph{(\bibinfo{series}{HILDA'17})}. \bibinfo{publisher}{Association for
  Computing Machinery}, \bibinfo{address}{New York, NY, USA}, Article
  \bibinfo{articleno}{7}, \bibinfo{numpages}{6}~pages.
\newblock
\showISBNx{9781450350297}
\urldef\tempurl%
\url{https://doi.org/10.1145/3077257.3077267}
\showDOI{\tempurl}


\bibitem[Miller et~al\mbox{.}(2018)]%
        {DBLP:journals/debu/MillerNZCPA18}
\bibfield{author}{\bibinfo{person}{Ren{\'{e}}e~J. Miller},
  \bibinfo{person}{Fatemeh Nargesian}, \bibinfo{person}{Erkang Zhu},
  \bibinfo{person}{Christina Christodoulakis}, \bibinfo{person}{Ken~Q. Pu},
  {and} \bibinfo{person}{Periklis Andritsos}.} \bibinfo{year}{2018}\natexlab{}.
\newblock \showarticletitle{Making Open Data Transparent: Data Discovery on
  Open Data}.
\newblock \bibinfo{journal}{\emph{{IEEE} Data Eng. Bull.}}
  \bibinfo{volume}{41}, \bibinfo{number}{2} (\bibinfo{year}{2018}),
  \bibinfo{pages}{59--70}.
\newblock
\urldef\tempurl%
\url{http://sites.computer.org/debull/A18june/p59.pdf}
\showURL{%
\tempurl}


\bibitem[Missier et~al\mbox{.}(2013)]%
        {missier2013w3c}
\bibfield{author}{\bibinfo{person}{Paolo Missier}, \bibinfo{person}{Khalid
  Belhajjame}, {and} \bibinfo{person}{James Cheney}.}
  \bibinfo{year}{2013}\natexlab{}.
\newblock \showarticletitle{The W3C PROV family of specifications for modelling
  provenance metadata}. In \bibinfo{booktitle}{\emph{EDBT '13}}.
  \bibinfo{pages}{773--776}.
\newblock


\bibitem[Nargesian et~al\mbox{.}(2019)]%
        {nargesian:2019:datalakemanagementtutorial}
\bibfield{author}{\bibinfo{person}{Fatemeh Nargesian}, \bibinfo{person}{Erkang
  Zhu}, \bibinfo{person}{Ren\'{e}e~J. Miller}, \bibinfo{person}{Ken~Q. Pu},
  {and} \bibinfo{person}{Patricia~C. Arocena}.}
  \bibinfo{year}{2019}\natexlab{}.
\newblock \showarticletitle{Data Lake Management: Challenges and
  Opportunities}.
\newblock \bibinfo{journal}{\emph{Proc. VLDB Endow.}} \bibinfo{volume}{12},
  \bibinfo{number}{12} (\bibinfo{date}{Aug.} \bibinfo{year}{2019}),
  \bibinfo{pages}{1986–1989}.
\newblock
\showISSN{2150-8097}
\urldef\tempurl%
\url{https://doi.org/10.14778/3352063.3352116}
\showDOI{\tempurl}


\bibitem[Nargesian et~al\mbox{.}(2018)]%
        {10.14778/3192965.3192973}
\bibfield{author}{\bibinfo{person}{Fatemeh Nargesian}, \bibinfo{person}{Erkang
  Zhu}, \bibinfo{person}{Ken~Q. Pu}, {and} \bibinfo{person}{Ren\'{e}e~J.
  Miller}.} \bibinfo{year}{2018}\natexlab{}.
\newblock \showarticletitle{Table Union Search on Open Data}.
\newblock \bibinfo{journal}{\emph{Proc. VLDB Endow.}} \bibinfo{volume}{11},
  \bibinfo{number}{7} (\bibinfo{date}{mar} \bibinfo{year}{2018}),
  \bibinfo{pages}{813–825}.
\newblock
\showISSN{2150-8097}
\urldef\tempurl%
\url{https://doi.org/10.14778/3192965.3192973}
\showDOI{\tempurl}


\bibitem[Naumann and Herschel(2010)]%
        {10.5555/1841211}
\bibfield{author}{\bibinfo{person}{Felix Naumann} {and}
  \bibinfo{person}{Melanie Herschel}.} \bibinfo{year}{2010}\natexlab{}.
\newblock \bibinfo{booktitle}{\emph{An Introduction to Duplicate Detection}}.
\newblock \bibinfo{publisher}{Morgan and Claypool Publishers}.
\newblock
\showISBNx{1608452204}


\bibitem[Niu et~al\mbox{.}(2017)]%
        {xing:2017:gprom}
\bibfield{author}{\bibinfo{person}{Xing Niu}, \bibinfo{person}{Raghav Kapoor},
  \bibinfo{person}{Boris Glavic}, \bibinfo{person}{Dieter Gawlick},
  \bibinfo{person}{Zhen~Hua Liu}, \bibinfo{person}{Vasudha Krishnaswamy}, {and}
  \bibinfo{person}{Venkatesh Radhakrishnan}.} \bibinfo{year}{2017}\natexlab{}.
\newblock \showarticletitle{Provenance-aware Query Optimization}. In
  \bibinfo{booktitle}{\emph{ICDE}}.
\newblock


\bibitem[Olston and Sarma(2011)]%
        {olston2011ibis}
\bibfield{author}{\bibinfo{person}{Christopher Olston} {and}
  \bibinfo{person}{Anish~Das Sarma}.} \bibinfo{year}{2011}\natexlab{}.
\newblock \showarticletitle{Ibis: A Provenance Manager for Multi-Layer
  Systems.}. In \bibinfo{booktitle}{\emph{CIDR}}. \bibinfo{pages}{152--159}.
\newblock


\bibitem[precisely(2022)]%
        {url:precisely}
precisely \bibinfo{year}{2022}\natexlab{}.
\newblock \bibinfo{title}{{Precisely}}.
\newblock \bibinfo{howpublished}{{\small\url{https://www.precisely.com}}}.
\newblock


\bibitem[Psallidas and Wu(2018)]%
        {psallidas2018smoke}
\bibfield{author}{\bibinfo{person}{Fotis Psallidas} {and}
  \bibinfo{person}{Eugene Wu}.} \bibinfo{year}{2018}\natexlab{}.
\newblock \showarticletitle{Smoke: Fine-grained lineage at interactive speed}.
\newblock \bibinfo{journal}{\emph{arXiv preprint arXiv:1801.07237}}
  (\bibinfo{year}{2018}).
\newblock


\bibitem[purview(2022)]%
        {url:purview}
purview \bibinfo{year}{2022}\natexlab{}.
\newblock \bibinfo{title}{{Microsoft Purview}}.
\newblock
  \bibinfo{howpublished}{{\small\url{https://azure.microsoft.com/en-us/services/purview}}}.
\newblock


\bibitem[Rupprecht et~al\mbox{.}(2020)]%
        {10.14778/3415478.3415556}
\bibfield{author}{\bibinfo{person}{Lukas Rupprecht}, \bibinfo{person}{James~C.
  Davis}, \bibinfo{person}{Constantine Arnold}, \bibinfo{person}{Yaniv Gur},
  {and} \bibinfo{person}{Deepavali Bhagwat}.} \bibinfo{year}{2020}\natexlab{}.
\newblock \showarticletitle{Improving Reproducibility of Data Science Pipelines
  through Transparent Provenance Capture}.
\newblock \bibinfo{journal}{\emph{Proc. VLDB Endow.}} \bibinfo{volume}{13},
  \bibinfo{number}{12} (\bibinfo{date}{aug} \bibinfo{year}{2020}),
  \bibinfo{pages}{3354–3368}.
\newblock
\showISSN{2150-8097}
\urldef\tempurl%
\url{https://doi.org/10.14778/3415478.3415556}
\showDOI{\tempurl}


\bibitem[Scherbaum et~al\mbox{.}(2018)]%
        {scherbaum2018spline}
\bibfield{author}{\bibinfo{person}{Jan Scherbaum}, \bibinfo{person}{Marek
  Novotny}, {and} \bibinfo{person}{Oleksandr Vayda}.}
  \bibinfo{year}{2018}\natexlab{}.
\newblock \showarticletitle{Spline: Spark lineage, not only for the banking
  industry}. In \bibinfo{booktitle}{\emph{2018 IEEE International Conference on
  Big Data and Smart Computing (BigComp)}}. IEEE, \bibinfo{pages}{495--498}.
\newblock


\bibitem[semanticweb(2022)]%
        {url:semanticweb}
semanticweb \bibinfo{year}{2022}\natexlab{}.
\newblock \bibinfo{title}{{Semantic Web Company}}.
\newblock \bibinfo{howpublished}{{\small\url{https://semantic-web.com}}}.
\newblock


\bibitem[Shneiderman(1996)]%
        {shneiderman1996eyes}
\bibfield{author}{\bibinfo{person}{Ben Shneiderman}.}
  \bibinfo{year}{1996}\natexlab{}.
\newblock \showarticletitle{The eyes have it: A task by data type taxonomy for
  information visualizations}. In \bibinfo{booktitle}{\emph{Symposium on Visual
  Languages}}. \bibinfo{pages}{336--343}.
\newblock


\bibitem[smartlogic(2022)]%
        {url:smartlogic}
smartlogic \bibinfo{year}{2022}\natexlab{}.
\newblock \bibinfo{title}{{Smartlogic}}.
\newblock \bibinfo{howpublished}{{\small\url{https://www.smartlogic.com}}}.
\newblock


\bibitem[sqldb(2022)]%
        {url:sqlmi}
sqldb \bibinfo{year}{2022}\natexlab{}.
\newblock \bibinfo{title}{{Azure SQL Managed Instance}}.
\newblock
  \bibinfo{howpublished}{{\small\url{https://azure.microsoft.com/en-us/products/azure-sql/managed-instance}}}.
\newblock


\bibitem[syniti(2022)]%
        {url:syniti}
syniti \bibinfo{year}{2022}\natexlab{}.
\newblock \bibinfo{title}{{Syniti}}.
\newblock \bibinfo{howpublished}{{\small\url{https://www.syniti.com}}}.
\newblock


\bibitem[Tang et~al\mbox{.}(2019)]%
        {tang:2019:sac}
\bibfield{author}{\bibinfo{person}{Mingjie Tang}, \bibinfo{person}{Saisai
  Shao}, \bibinfo{person}{Weiqing Yang}, \bibinfo{person}{Yanbo Liang},
  \bibinfo{person}{Yongyang Yu}, \bibinfo{person}{Bikas Saha}, {and}
  \bibinfo{person}{Dongjoon Hyun}.} \bibinfo{year}{2019}\natexlab{}.
\newblock \showarticletitle{SAC: A System for Big Data Lineage Tracking}. In
  \bibinfo{booktitle}{\emph{2019 IEEE 35th International Conference on Data
  Engineering (ICDE)}}. \bibinfo{pages}{1964--1967}.
\newblock
\urldef\tempurl%
\url{https://doi.org/10.1109/ICDE.2019.00215}
\showDOI{\tempurl}


\bibitem[Widom(2005)]%
        {widom:2005:trio}
\bibfield{author}{\bibinfo{person}{Jennifer Widom}.}
  \bibinfo{year}{2005}\natexlab{}.
\newblock \showarticletitle{Trio: a system for integrated management of data,
  accuracy, and lineage}. In \bibinfo{booktitle}{\emph{CIDR}}.
\newblock


\bibitem[Woodruff and Stonebraker(1997)]%
        {woodruff1997supporting}
\bibfield{author}{\bibinfo{person}{Allison Woodruff} {and}
  \bibinfo{person}{Michael Stonebraker}.} \bibinfo{year}{1997}\natexlab{}.
\newblock \showarticletitle{Supporting fine-grained data lineage in a database
  visualization environment}. In \bibinfo{booktitle}{\emph{ICDE}}.
\newblock


\bibitem[xelite(2022)]%
        {url:xelite}
xelite \bibinfo{year}{2022}\natexlab{}.
\newblock \bibinfo{title}{{Microsoft.SqlServer.XEvent.XELite}}.
\newblock \bibinfo{howpublished}{{\small
  \url{https://www.nuget.org/packages/Microsoft.SqlServer.XEvent.XELite/}}}.
\newblock


\bibitem[xevents(2019)]%
        {url:xevents}
xevents \bibinfo{year}{2019}\natexlab{}.
\newblock \bibinfo{title}{{Extended Events Overview}}.
\newblock \bibinfo{howpublished}{{\small
  \url{https://docs.microsoft.com/en-us/sql/relational-databases/extended-events/extended-events}}}.
\newblock


\bibitem[Yakout et~al\mbox{.}(2011)]%
        {10.14778/1952376.1952378}
\bibfield{author}{\bibinfo{person}{Mohamed Yakout}, \bibinfo{person}{Ahmed~K.
  Elmagarmid}, \bibinfo{person}{Jennifer Neville}, \bibinfo{person}{Mourad
  Ouzzani}, {and} \bibinfo{person}{Ihab~F. Ilyas}.}
  \bibinfo{year}{2011}\natexlab{}.
\newblock \showarticletitle{Guided Data Repair}.
\newblock \bibinfo{journal}{\emph{Proc. VLDB Endow.}} \bibinfo{volume}{4},
  \bibinfo{number}{5} (\bibinfo{date}{feb} \bibinfo{year}{2011}),
  \bibinfo{pages}{279–289}.
\newblock
\showISSN{2150-8097}
\urldef\tempurl%
\url{https://doi.org/10.14778/1952376.1952378}
\showDOI{\tempurl}


\bibitem[Zhu et~al\mbox{.}(2016)]%
        {10.14778/2994509.2994534}
\bibfield{author}{\bibinfo{person}{Erkang Zhu}, \bibinfo{person}{Fatemeh
  Nargesian}, \bibinfo{person}{Ken~Q. Pu}, {and} \bibinfo{person}{Ren\'{e}e~J.
  Miller}.} \bibinfo{year}{2016}\natexlab{}.
\newblock \showarticletitle{LSH Ensemble: Internet-Scale Domain Search}.
\newblock \bibinfo{journal}{\emph{Proc. VLDB Endow.}} \bibinfo{volume}{9},
  \bibinfo{number}{12} (\bibinfo{date}{aug} \bibinfo{year}{2016}),
  \bibinfo{pages}{1185–1196}.
\newblock
\showISSN{2150-8097}
\urldef\tempurl%
\url{https://doi.org/10.14778/2994509.2994534}
\showDOI{\tempurl}


\end{thebibliography}

\end{document}